\documentclass[aps,prx,twocolumn,superscriptaddress,showpacs,longbibliography]{revtex4-1}

\usepackage{amsmath}
\usepackage{graphicx}
\usepackage{amsfonts}
\usepackage{verbatim}
\usepackage{dsfont}
\usepackage{color}
\usepackage{hyperref}
\hypersetup{colorlinks = true, urlcolor = blue, linkcolor = blue, citecolor = blue}

\begin{document}
\title{Full proximity treatment of topological superconductors 
in Josephson-junction architectures}

\author{F. Setiawan}\email{setiawan@uchicago.edu}
\affiliation{James Franck Institute, University of Chicago, Chicago, Illinois 60637, USA}
\author{Chien-Te Wu}
\affiliation{Department of Electrophysics, National Chiao Tung University, Hsinchu 30010, Taiwan, Republic of China}
\affiliation{Physics Division, National Center for Theoretical Sciences, Hsinchu 30010, Taiwan, Republic of China}
\affiliation{James Franck Institute, University of Chicago, Chicago, Illinois 60637, USA}
\author{K. Levin}
\affiliation{James Franck Institute, University of Chicago, Chicago, Illinois 60637, USA}

\begin{abstract}

Experiments on planar Josephson-junction architectures have recently been shown to provide an alternative way of creating topological superconductors hosting accessible Majorana modes. These zero-energy modes can be found at the ends of a one-dimensional channel in the junction of a two-dimensional electron gas (2DEG) proximitized by two spatially separated superconductors. The channel, which is below the break between the superconductors, is not in direct contact with the superconducting leads, so that proximity coupling is expected to be weaker and less well controlled than in the simple nanowire configuration widely discussed in the literature. This provides a strong incentive for this paper which investigates the nature of proximitization in these Josephson junction architectures. At a microscopic level we demonstrate how and when it can lead to topological phases. We do so by going beyond simple tunneling models through solving self-consistently the Bogoliubov-de Gennes equations of a heterostructure multicomponent system involving two spatially separated $s$-wave superconductors in contact with a normal Rashba spin-orbit-coupled 2DEG. Importantly, within our self-consistent theory we present ways of maximizing the proximity-induced superconducting gap by studying the effect of the Rashba spin-orbit coupling, chemical potential mismatch between the superconductor and 2DEG, and sample geometry on the gap. Finally, we note (as in experiment) a Fulde-Ferrell-Larkin-Ovchinnikov phase is also found to appear in the 2DEG channel, albeit under circumstances which are not ideal for the topological superconducting phase.

\end{abstract}
\maketitle
\vskip5mm

\section{Introduction} 

There has been much excitement in the literature over the possibility of observing one-dimensional (1D) topological superconductivity
which involves a single 1D wire~\cite{Lutchyn2010Majorana,Oreg2010Helical}
leading to accessible Majorana zero modes.
Because of fluctuation effects in low
dimensions, there can be no
intrinsic superconductivity so that the focus is on proximitized
superconductors. 
Studies of these wires and their applications towards
quantum computation have led to a very extensive literature~\cite{alicea2012new,leijnse2012_introduction,stanescu2013majorana,Beenakker2013Search,Elliott2015Colloquium,sarma2015majorana,aguado2017majorana,lutchyn2018majorana}.
In a broad sense, there are two general configurations for proximitized 1D
topological superconductors. These are associated with
``nanowires" in direct contact
with superconducting hosts as well as the recently proposed planar Josephson junction~\cite{Pientka2017Topological,Hell2017Two}.
The latter contains a proximitized 1D channel in the two-dimensional electron gas (2DEG)
just below the break between the two superconductors. 
This configuration is less widely studied, 
but there is evidence based on zero-bias conductance peaks~\cite{Krishnendu2001Midgap,Law2009Majorana,Flensberg2010Tunneling,Elsa2012Transport}, as in the simple nanowires~\cite{Mourik2012Signatures,Rokhinson2012fractional,Deng2012Anomalous,Das2012Zero,Churchill2013Superconductor,Finck2013Anomalous,albrecht2016exponential,gul2018ballistic,chen2017experimental,deng2016majorana,Suominen2017Zero,Nichele2017Scaling,zhang2018quantized,zhang2017ballistic,Sestoft2018Engineering,Deng2018Nonlocality,laroche2017observation,van2018observation,de2018electric,grivnin2018concomitant,vaitiekenas2018flux}, that topological superconductivity has been
experimentally observed~\cite{Ren2018Topological,Antonio2018Evidence}.

\begin{figure} 
\includegraphics[width=\linewidth]{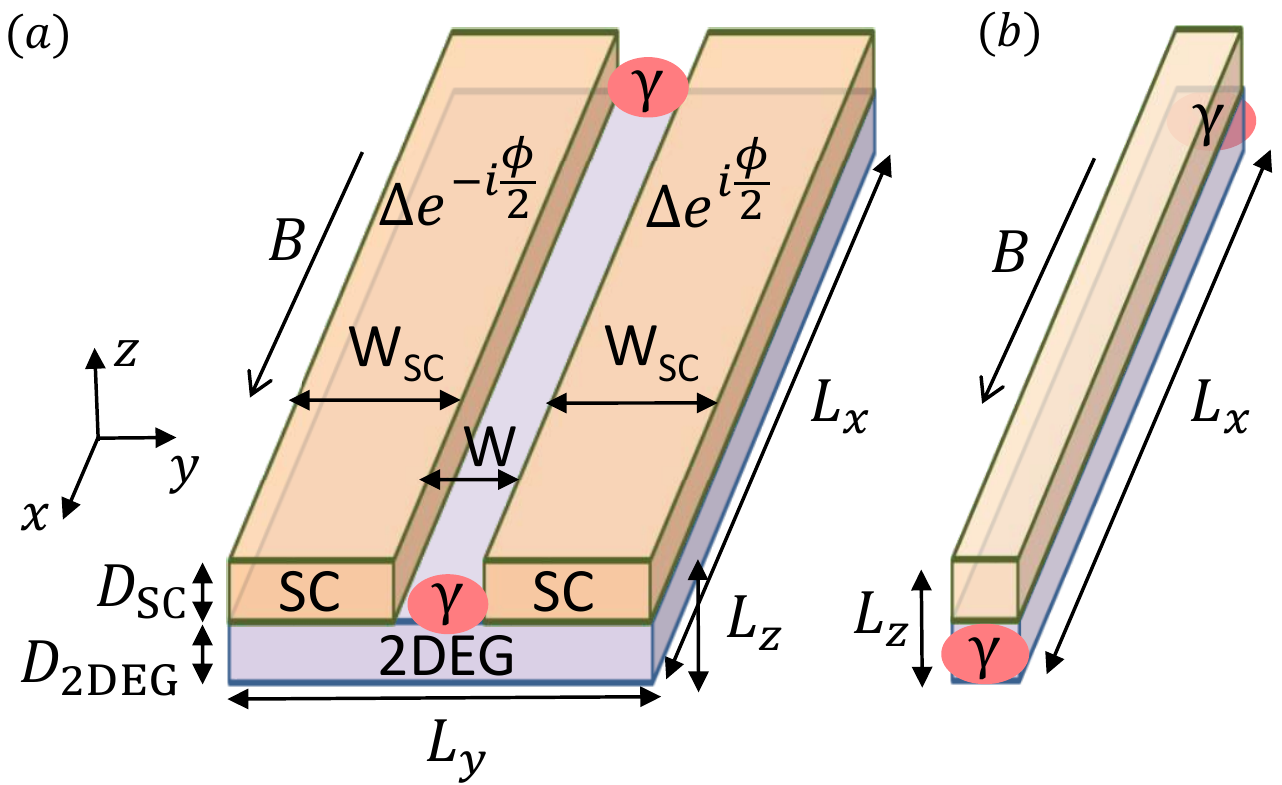}
\caption{(a) Schematic diagram of a 2DEG in proximity to two spatially separated superconducting leads which form a Josephson junction. By tuning the strength of either the applied in-plane magnetic field $B$ or the phase difference $\phi$ between the two superconductors, the system can be tuned into the topological superconducting phase which hosts Majorana zero modes ($\gamma$) at the end of the junction. (b) Schematic diagram of a nanowire proximitized by a superconductor. The system becomes a topological superconductor, which hosts Majorana zero modes ($\gamma$) at the end of the nanowire, when the strength of the magnetic field $B$ is above a certain critical value. }
\label{fig:fig1}
\end{figure}

Indeed, the planar junctions have a notable strength relative to the nanowires. The phase difference between the two superconductors provides an alternative knob (beyond the Zeeman field) to tune the system into the topological 
phase~\cite{Pientka2017Topological,Hell2017Two}.  In ideal (i.e., transparent) systems, when the superconducting phase difference is $\phi = \pi$, the topological phase  
can be achieved for rather small Zeeman fields. 
However, compared to the proximitized nanowire, the planar Josephson junction architecture is associated with
weaker and less well-controlled proximitization, as the 1D channel in the junction is
not in direct contact with the host superconductors.

This leads to the central goal of this paper which is to quantify this 
somewhat indirect form of proximitization
and to optimize its effectiveness.
We focus on a well-studied substrate:
the 2DEG which has moderately strong
Rashba spin-orbit coupling (SOC).
Our calculations go beyond the simple tunneling
models \cite{McMillan1968tunneling,McMillan1968theory,Sau2010robustness,Stanescu2010Proximity,Potter2011engineering} of the proximity effect by solving the full Bogoliubov-de Gennes (BdG) equations of a multicomponent system with 
self consistency~\footnote{Although not all of the results presented in this paper are fully self-consistent, we have checked in a few cases that increasing the number of iterations only very slightly modifies our results.}.
In our full proximity model, the host superconductors are treated as a participating component rather than as a passive source of Cooper
pairing.
The effectiveness of proximitization is quantified via the
strength of the induced pairing amplitude, $\Delta^{\mathrm{prox}}$.
Maximizing this pairing amplitude is the goal as it is associated with a
large gap in the dispersion. This, in turn, leads to more localized and thus more
stable Majorana modes.
In this paper we characterize
the deleterious effects on $\Delta^{\mathrm{prox}}$ which can come from
any of the following: SOC, enhanced substrate thickness, enhanced channel width, and chemical
potential differences (between the host superconductors
and the 2DEG).
Importantly, our findings which are obtained using a fully self-consistent theory, can provide guidance in 
determining the optimal range of experimental parameters for the topological protection of Majorana modes.

While not essential to the topological superconductivity, a relevant
complement to
these studies relates to a very elusive state of matter, the  Fulde-Ferrell-Larkin-Ovchinnikov (FFLO)~\cite{Fulde1964Superconductivity,Larkin1965Nonuniform} phase
which we also observe in these planar junctions.
This appears to be consistent with recent experiments
which have reported that this otherwise rare phase of superconductivity is
realized in proximitized superconductors
~\cite{hart2017controlled,chen2018finite}.
For the situation here, it can 
be viewed as
arising from a ``second-order
proximitization" process. We trace its origin
to the fact that
the channel makes little direct contact with the superconductors,
unlike the rest of the proximitized 2DEG. Thus,
in this region of the junction, the pairing amplitude is 
reduced and the effective small pairing gap is
freer to oscillate in response to an
applied Zeeman field.
We finally note that this FFLO phase
is most apparent in relatively wide junctions where the gap is smaller and it is
thus unfavorable for stabilizing a topological phase.

The two generic types of proximitized 1D topological superconductors are illustrated in
Fig.~\ref{fig:fig1}. The Majorana zero-modes (indicated by $\gamma$) appear at the ends of the junction where they are
most easily manipulated. 
In structures as shown in Fig.~\ref{fig:fig1}(a), the substrate is 
a Rashba spin-orbit-coupled 2DEG. Figure~\ref{fig:fig1}(b) shows a more successful variant of these
hybrid structures which involve semiconducting
nanowires (although chains of magnetic atoms~\cite{Yaz1,Yaz2,Yaz3,Yaz4,Yaz5,Brydon2015topological,hui2015majorana} and topological insulators~\cite{Fu2008Superconducting,Xu2015Experimental,HaoHua2016Majorana}
have also been considered). 

One should appreciate that to design topological superconductors 
without proximitization, say by doping a topological insulator~\cite{Fu2010odd,Satoshi2011Topological}, there
is less control in
engineering the
appropriate combination of SOC,
Zeeman field, and band structure in the presence of sufficiently
strong pairing attraction.
The existence of these intrinsic topological superconductors is still controversial~\cite{Haibing2013Absence}
so that, currently, 
proximity-induced superconductivity appears to be
an essential tool.
Because it is so essential it is imperative to understand it 
better, not just in the immediate interface, which has been studied \cite{Potter2011engineering,Stanescu2010Proximity,Sau2010robustness},
but well into the depth of a hypothesized topological
superconductor~\cite{Chiu2016Induced,Stanescu2013Superconducting}.

\subsection{Overview and Outline}
It is useful to quantitatively characterize the Josephson-junction-based topological superconductors
we consider here
in terms of the size of the energy gap, $E_{\mathrm{gap}}$, associated with the proximitized 2DEG.
The quantity $E_{\mathrm{gap}}$ depends on the junction geometry and
materials parameters. It varies with
the junction thickness, the strip
width, the SOC and chemical potential difference between the host
superconductors and the 2DEG. 
Equally important is its dependence 
on the external parameters which control topological phases: the Josephson-junction
phase difference $\phi$ and the
Zeeman field $E_Z$.
This
field enters in two different ways; it affects the 
gap opening and closing processes associated with topological phase transitions
in a Josephson junction.
It also affects
the coupling at each separate interface
between the host superconductor and the 2DEG substrate. Increasing $E_Z$ in the 2DEG inhibits proximitization. 

It is convenient, then, to isolate these processes by writing
\begin{equation}
E_{\mathrm{gap}}\equiv \Delta^{\mathrm{prox}}|_{(E_Z= \phi = 0)}~~ f (E_Z, \phi).
\end{equation}
This states that the energy gap in
the presence of Zeeman and superconducting phase difference, $E_{\mathrm{gap}}$, depends directly on a proximity-induced gap
$\Delta^{\mathrm{prox}}$, (which is deduced in the absence of any Zeeman field, $E_Z$ or phase bias
$\phi$),
times a 
multiplicative function, $f(E_Z,\phi)$, which represents (dominantly) the topological
characteristics of the junction.

In the topological region, the parameter $E_{\mathrm{gap}}$ is, thus, a crucial parameter,
as its inverse characterizes the Majorana localization length.
The smaller this length, the more localized are the Majorana modes.
The localization of the Majoranas is, then, optimized when the proximity gap $\Delta^{\mathrm{prox}}$ is maximal.
Understanding this is one of the central contributions of our paper.

We now present a brief outline. Section~\ref{sec:model} of the paper discusses the theoretical model, i.e., the Hamiltonian of the planar Josephson junction. In Sec.~\ref{sec:BdG}, we give a discussion of the self-consistent BdG approach used to solve for the energy dispersion and proximity-induced gap.
In Sec.~\ref{sec:tunneling} we study a simple tunneling model of the superconducting proximity effect in which the junction is converted to
a lower dimension by integrating out the host superconductors.
Section~\ref{sec:proximitygap}
focuses on numerical results from
our full-proximity model for the proximity gap $\Delta^{\mathrm{prox}}$ where $\Delta^{\mathrm{prox}}$ is the spectral gap calculated for junctions in the absence of Zeeman field and superconducting phase difference. Here we separately
discuss the role of SOC, chemical potential mismatch, and 2DEG thickness on $\Delta^{\mathrm{prox}}$. 
 The symmetry class of the planar Josephson junction is addressed in Sec.~\ref{sec:symmetry}. In Sec.~\ref{sec:topological}  
we present the topological phase diagram as a function of in-plane Zeeman field and superconducting phase bias for different chemical potential mismatch. We further show the evolution of the energy spectrum across the topological phase transition.
Section~\ref{sec:FFLO} presents a brief discussion of how FFLO superconducting phase
is established, in the presence of an in-plane Zeeman
field along the junction.
More details of this elusive FFLO phase are presented in Appendix~\ref{sec:appendix}. Finally, we 
summarize our conclusions in Sec.~\ref{sec:conclusion}.

\section{Theoretical Model}\label{sec:model}
We consider a Josephson junction made from a Rashba spin-orbit-coupled 2DEG in contact with two spatially separated superconductors and subjected to an in-plane magnetic field along the junction as shown in Fig.~\ref{fig:fig1}(a). This system was proposed recently~\cite{Pientka2017Topological,Hell2017Two} as a new platform to realize topological superconductors. In this setup, the transition between the trivial and topological phases can be tuned by varying either
the applied in-plane magnetic field $B$ along the junction or the phase difference $\phi$ between the two superconductors. 
In an ideal situation, the interplay between these two independent knobs enables a lower critical field for the topological transition to be achieved  when the superconducting phase difference is tuned near $\phi = \pi$. This Zeeman- and phase-tunable topological transition was demonstrated in recent experiments carried out by two independent groups~\cite{Ren2018Topological,Antonio2018Evidence}.

\subsection{Hamiltonian}
We begin by writing down the ``normal" component (in the absence of superconducting pairing) of the Hamiltonian as
\begin{align}\label{eq:Hnormal}
H & =&\int d^{3}\boldsymbol{r}\sum_{\sigma\sigma'}\psi_{\sigma}^{\dagger}(\boldsymbol{r})\left[\left(\frac{\boldsymbol{P}^{2}}{2m^*}-\mu(\boldsymbol{r})\right)\sigma_{0}+E_{Z}(\boldsymbol{r})\sigma_{x}\right. \nonumber \\
&&+\left.\alpha\left(\boldsymbol{r}\right)\left(P_{x}\sigma_{y}-P_y\sigma_{x}\right)\right]\psi_{\sigma'}\left(\boldsymbol{r}\right),
\end{align}
where $\psi_{\sigma}$ ($\psi_{\sigma}^{\dagger}$) is the annihilation (creation) operator of an electron with spin $\sigma = \uparrow,\downarrow$. In Eq.~({\ref{eq:Hnormal}}), $\sigma_0$ is the identity matrix and
$\boldsymbol{\sigma}\equiv(\sigma_x,\sigma_y,\sigma_z)$ are the Pauli matrices acting on the spin degree of freedom. Here, $\boldsymbol{P}$ represents the 
real space momentum
operator, $m^*$ is the effective electron mass, and $\mu$ is the chemical potential.
The chemical potentials are taken to be
\begin{equation}\label{eq:muyz}
\mu(\boldsymbol{r}) = \begin{cases}
\mu_{\mathrm{S}} & \text{for $W/2 < |y| < W_{\mathrm{SC}}+W/2$}\\
& \text{\hspace{0.5 cm} and $ D_{\mathrm{2DEG}}< z < D_{\mathrm{2DEG}}+D_{\mathrm{SC}} $},\\
\mu_{\mathrm{2DEG}} & \text{for $|y| < W_{\mathrm{SC}}+W/2$}\\
&   \text{\hspace{0.5 cm} and $0 < z < D_{\mathrm{2DEG}}$},
\end{cases}
\end{equation}
where $\mu_{\mathrm{S}}$ and $\mu_{\mathrm{2DEG}}$ are the chemical potentials of the superconductor and 2DEG, respectively. Throughout this paper, we work in units where $\hbar = 1$, $\mu_{\mathrm{2DEG}} = 1$, and $2m^* = 1$ which gives the Fermi momentum of the 2DEG, $k_F = 1$. The widths of the superconductors and the junction (along the $y$ direction) are denoted by $W_{\mathrm{SC}}$ and $W$, respectively [see Fig.~\ref{fig:fig1}(a)].  In this paper we consider the width of the superconducting leads $W_{\mathrm{SC}} >  \xi$, where $\xi$ is the superconducting coherence length. We further denote the thicknesses of the superconductors and the 2DEG by $D_{\mathrm{SC}}$ and $D_{\mathrm{2DEG}}$, respectively.   Note that for numerical simplicity, we introduce an insulator in between the superconductors with the same thickness as the superconductor above the 2DEG. Its chemical potential is taken to be very negative ($\mu_I = -5$), so
that it behaves essentially as a vacuum. 

The Zeeman energy $E_Z(\boldsymbol{r}) = \widetilde{g}(\boldsymbol{r})\mu_B B/2$ 
is due to the applied in-plane magnetic field $B$ along 
the junction ($x$ direction) with $\widetilde{g}$ being the Lande $g$ factor and $\mu_B$ being the Bohr magneton. Except when indicated otherwise,
the Zeeman energy $E_Z({\boldsymbol{r}})$ is assumed to be zero in the host
superconductor and insulator but taken to be constant throughout the 2DEG ($E_{Z,L} = E_{Z,J} = E_Z$, where $E_{Z,L}$ is the Zeeman energy of the 2DEG directly below the superconducting leads and $E_{Z,J}$ is the Zeeman energy of the 2DEG in the junction). We justify this assumption by noting that the Lande $g$ factor for the superconductor ($\widetilde{g}\sim 2$ for Al) is much smaller than the Lande $g$ factor for the semiconductor ($\widetilde{g}\sim 15$ for InAs)~\cite{winkler2003spin,Mikkelsen2018Hybridization,Effects2018Antipov}. 

An important
parameter which appears throughout this paper is $\alpha$ which characterizes the strength of the SOC in the 2DEG. 
The SOC strength is zero in the superconductors and insulator but finite in the 2DEG, i.e.,
\begin{equation}\label{eq:alpha}
\alpha(\boldsymbol{r}) = \begin{cases}
0 & \text{for $ D_{\mathrm{2DEG}}< z < D_{\mathrm{2DEG}}+D_{\mathrm{SC}} $},\\
\alpha & \text{for $ 0< z < D_{\mathrm{2DEG}}$}.
\end{cases}
\end{equation} This
is a realistic representation~\cite{Sticlet2017Robustness,Stanescu2017Proximity} of the 
well-studied situation of a spin-orbit-coupled semiconductor proximitized by an $s$-wave superconductor.

So far we have described a noninteracting system. Now, let us include the superconducting pairing term in the Hamiltonian, which is given by
\begin{equation}
\sum_{\sigma\sigma^\prime}(i\sigma_y)_{\sigma\sigma^\prime}\Delta(\boldsymbol{r})\psi_\sigma^\dagger(\boldsymbol{r})\psi_{\sigma^\prime}^\dagger(\boldsymbol{r})+\rm{H.c.}
\end{equation}
We assume that the system is translationally invariant along the $x$
direction and finite in both $y$ and $z$ directions. Because the system is translationally invariant along the $x$ direction, we can write the Hamiltonian in the Nambu basis $\Psi_{k_x}(y,z) = \left[\psi_{k_x\uparrow}(y,z), \psi_{k_x\downarrow}(y,z), \psi^\dagger_{k_x\downarrow}(y,z), -\psi^\dagger_{k_x\uparrow}(y,z)\right]^T$ as
\begin{align}
H &= \frac{1}{2}\int dk_x \int dz \int dy \Psi_{k_x}^\dagger(y,z) \mathcal{H}_{k_x}(y,z) \Psi_{k_x}(y,z),
\label{kxhamiltonian}
\end{align}
where the BdG Hamiltonian is given by
\begin{align}\label{eq:BdG}
\mathcal{H}_{k_x}(y,z) = &\left[k_x^2 - \partial_y^2 - \partial_z^2 - \mu(y,z)\right] \tau_z \nonumber\\
& + \alpha (z)( k_x \sigma_y + i \partial_y \sigma_x) \tau_z + E_Z(y,z)\sigma_x\nonumber\\
& + \Delta(y,z)\tau_+ + \Delta^*(y,z)\tau_-,
\end{align}
Here the Pauli matrices $\boldsymbol{\sigma}$ and $\boldsymbol{\tau}$ act in the spin and particle-hole subspace, respectively, with $\tau_{\pm} = (\tau_x\pm i\tau_y)/2$.
The superconducting pairing potential, $\Delta(y,z)$, arises microscopically from the attractive interactions which are only present
in the host superconductors:
\begin{equation}\label{eq:deltar}
\Delta (y,z) \equiv  g(y,z) F(y,z),
\end{equation}
where  $g(y,z)$ is the coupling function within the parent superconductors:
\begin{equation}\label{eq:gr}
g(y,z) = \begin{cases}
g_0e^{-i\phi/2} & \text{for $-(W_{\mathrm{SC}}+W/2)  < y < -W/2$}\\
& \text{\hspace{0.1 cm} and $ D_{\mathrm{2DEG}}< z < D_{\mathrm{2DEG}}+D_{\mathrm{SC}} $},\\
g_0e^{i\phi/2} & \text{for $ W/2 < y < W_{\mathrm{SC}}+W/2$}\\
& \text{\hspace{0.1 cm} and $ D_{\mathrm{2DEG}}< z < D_{\mathrm{2DEG}}+D_{\mathrm{SC}} $},\\
0 & \text{otherwise}.
\end{cases}
\end{equation}
Here, $g_0$ is the attractive coupling constant, and $\phi$ is the phase difference between the two superconductors.
Applying a Bogoliubov transformation, $\psi_{k_x\sigma}=\sum_n\left[u_{nk_x\sigma}\gamma_{nk_x}+v_{nk_x\sigma}^\ast\gamma_{nk_x}^\dagger\right]$~\cite{halterman,halterman2}, where $\gamma_{nk_x} (\gamma_{nk_x}^\dagger)$ is the Bogoliubov quasiparticle annihilation (creation) operator at an energy $E_n$, we then obtain the pair amplitude
\begin{align}\label{eq:Deltaself}
F(y,z) &=  \langle\psi_{\uparrow}(y,z) \psi_{\downarrow}(y,z)\rangle 
 \nonumber\\ 
&=  \int dk_x \sum_{E_{nk_x}<\omega_D} \left[u_{nk_x\uparrow}v_{nk_x\downarrow}^*-u_{nk_x\downarrow}v_{nk_x\uparrow}^*\right]\nonumber\\
&\hspace{4 cm}\times\tanh\left(\frac{E_{nk_x}}{2T}\right),
\end{align}
with $T$ being the temperature. The Debye frequency $\omega_D$ provides an energy cutoff in Eq.~\eqref{eq:Deltaself}. Note that, through the proximity effect,
the pair amplitude $F(y,z)$  in the 2DEG is nonzero even though there is a vanishing order parameter, $\Delta=0$, reflecting the fact that $g(y,z)=0$ there. 
The superconducting pairing potential $\Delta(y,z)$ is obtained by solving the BdG Hamiltonian self-consistently as explained in the next subsection. 

\section{Self-Consistent B\lowercase{d}G Equation}\label{sec:BdG}
We obtain the pair amplitude $F(y,z)$ [Eq.~\eqref{eq:Deltaself}] by numerically solving the BdG eigenvalue problem following the scheme
developed in Refs.~\cite{halterman,halterman2,Wu2017Majorana,Wu2018Quantum}. The scheme is based on the idea of diagonalizing the BdG Hamiltonian [Eq.~\eqref{eq:BdG}]. The resulting BdG equation reads
\begin{equation}
\label{bdgH}
\mathcal{H}_{k_x}(y,z) \Phi_{nk_x}(y,z) = E_{nk_x} \Phi_{nk_x}(y,z),
\end{equation}
where the wave function is given by
\begin{align}\label{eq:phin}
\Phi_{nk_x}(y,z) = \left(\begin{matrix} u_{nk_x\uparrow}(y,z) \\ u_{nk_x\downarrow}(y,z) \\ v_{nk_x\downarrow}(y,z) \\ -v_{nk_x\uparrow}(y,z)\end{matrix}\right),
\end{align}
with the boundary condition $\Phi_{nk_x}(y,z) = 0$ at $|y|>W_{\mathrm{SC}}+W/2$, $z <0$ and $z>D_{\mathrm{2DEG}}+D_{\mathrm{SC}}$ and subject to the self-consistency equation~[Eqs.~\eqref{eq:deltar}-~\eqref{eq:Deltaself}]. To this end,
we expand both the matrix elements and
the eigenfunctions in terms of a Fourier basis. Specifically, the quasiparticle ($u_{nk_x\sigma}$) and quasihole ($v_{nk_x\sigma}$) wavefunctions are given by
\begin{subequations}\label{eq:uv}
\begin{align}
u_{nk_x\sigma}(y,z)&=&\frac{2}{\sqrt{L_yL_z}}\sum_{pq}u_{nk_x\sigma}^{pq}\sin\left(\frac{p\pi y}{L_y}\right)\sin\left(\frac{q\pi z}{L_z}\right),\\
v_{nk_x\sigma}(y,z)&=&\frac{2}{\sqrt{L_yL_z}}\sum_{pq}v_{nk_x\sigma}^{pq}\sin\left(\frac{p\pi y}{L_y}\right)\sin\left(\frac{q\pi z}{L_z}\right).
\end{align}
\end{subequations}
For definiteness,  we set the smallest length scale to be
of the order of $1/k_F$ where $k_F  = \sqrt{\mu_{\mathrm{2DEG}}}$ is the Fermi momentum of the 2DEG. 

General matrix elements are similarly expanded in terms of the same Fourier series.
For example, we define the matrix elements of an operator $O$ to be
\begin{eqnarray}
O^{pqp'q'} & \equiv & \langle pq|O|p'q'\rangle \nonumber\\
& =&
\frac{4}{L_yL_z}\int_{0}^{L_y}\int_{0}^{L_z}dydz \sin\left(\frac{p\pi y}{L_y}\right)\sin\left(\frac{q\pi z}{L_z}\right)\nonumber
\\
& &
 \qquad \times O \sin\left(\frac{p'\pi y}{L_y}\right)\sin\left(\frac{q'\pi z}{L_z}\right).
\end{eqnarray}
In this way all terms in the BdG Hamiltonian can be expanded in this basis set.
What we have accomplished in this procedure is to
successfully transform a set of differential equations into an algebraic matrix eigenvalue
problem. 

Having recast the Hamiltonian in the basis given in Eq.~\eqref{eq:uv}, we then solve for the pair amplitude using Eqs.~\eqref{eq:deltar}-\eqref{eq:Deltaself} from the wavefunction [Eq.~\eqref{eq:phin}] obtained by diagonalizing the Hamiltonian~[Eq.~\eqref{bdgH}]. The calculated pair amplitude is then used to get a new wavefunction . This self-consistent procedure is carried out repeatedly until convergence is reached.  The first iteration generally contains the central physics. Because of the numerical complexity of the full-proximity model and the many parameter sets we address, in many plots we restrict ourselves to the first iteration; in test cases we have confirmed that higher iterations introduce changes in the solution of only a few percent. Throughout this paper, the pair amplitude $F(y,z)$ is calculated by setting the parent superconductor pair potential, $\Delta_0=0.3$,
Debye frequency $\omega_D=0.5$, and temperature $T=0$ in Eq.~\eqref{eq:Deltaself}. 

\section{Tunneling approximation to proximitization}\label{sec:tunneling}
The above more powerful procedure has not been widely applied; rather the literature focus has
been on an approximate treatment of proximitization. The approximate approach builds on earlier work
by 
McMillan~\cite{McMillan1968tunneling,McMillan1968theory}, who introduced
a perturbative treatment of a tunneling Hamiltonian for
a single NS junction which consists of a normal metal in proximity to a superconductor. This treatment was later extended by Refs.~\cite{Sau2010robustness,Stanescu2010Proximity,Potter2011engineering} to deal with a spin-orbit-coupled electron gas or a topological insulator in proximity with a superconductor. In this section we use
$N$ and $S$ to represent the 2DEG and superconductor, respectively; both are
considered to be sufficiently thin so that any spatial variations
within each can be ignored. The Hamiltonian for the SC/2DEG heterostructure can be written as 
\begin{equation}\label{eq:totH}
H = H_{S} + H_{N} + H_{T}. 
\end{equation}
Here, $H_{S,N}$ is the Hamiltonian of the superconductor ($S$) and 2DEG ($N$), respectively, and the tunneling Hamiltonian is given by
\begin{equation}
H_T = \sum_{\mathbf{k}_\parallel,k_\perp,\sigma} t (c_{S,(\mathbf{k}_{\parallel},k_{\perp}),\sigma}^{\dagger} c_{N,\mathbf{k}_{\parallel},\sigma}) + \mathrm{h.c.},
\end{equation}
where $c_{S/N,k,\sigma}$ is the annihilation operator in the $S$ or $N$ side of the interface for an electron with momentum $k$ and spin $\sigma =\left\uparrow\right./\left\downarrow\right.$.  This tunnel Hamiltonian $H_T$ conserves momentum $\mathbf{k}_{\parallel}$ parallel to the $NS$ interface but changes the transverse momentum $k_{\perp}$ perpendicular to the interface.

In this approach one derives the proximity-induced superconductivity by integrating out the superconducting term in Eq.~\eqref{eq:totH} and calculating the surface self-energy due to the electron tunneling between the 2DEG and superconductor. 

Assuming the density of states to be weakly dependent on energy, the surface self-energy can be calculated to be~\cite{Sau2010robustness,Stanescu2011Majorana}
\begin{align}\label{eq:sigman}
\Sigma_N (\omega) &= |t|^2 \nu(\varepsilon_{F_N}) \int d\varepsilon G_S(\varepsilon,\omega)\nonumber\\
& = -|t|^2 \nu(\varepsilon_{F_N}) \left[\frac{\omega\tau_0 + \Delta_0 \tau_x }{\sqrt{\Delta_S^2 -\omega^2}} + \zeta_N \tau_z\right],
\end{align}
where the density of states $\nu(\varepsilon_{F_N})$ is evaluated at the Fermi energy of the 2DEG and $\zeta_N$ is the proximity-induced shift in the chemical potential of the 2DEG. We can now incorporate this self-energy into the strong-coupling form~\cite{Nambu1960Quasi,eliashberg1960interactions,schrieffer2018theory} of the 2DEG Green's function, where we have
\begin{align}\label{eq:GN}
G_{N}(\boldsymbol{k},\omega) &= \frac{Z_{\Gamma_N}}{\omega - Z_{\Gamma_N}H_N -(1-Z_{\Gamma_N})\Delta_S\tau_x}.
\end{align}
Here,
\begin{align}
Z_{\Gamma_N}(\omega) = \left(1+\frac{\Gamma_N}{\sqrt{\Delta_S^2 -\omega^2}}\right)^{-1}
\end{align}
is the reduced quasiparticle weight due to the virtual propagation of electrons in the superconductor with  $\Gamma_N = |t|^2\nu(\varepsilon_{F,N})$ being the effective coupling between the 2DEG and superconductor. This quasiparticle weight can be viewed as the fraction
of time that a propagating electron spends on the superconducting side of the $NS$ interface. The proximity-induced superconducting pairing potential in the 2DEG is then given by
\begin{equation}\label{eq:DeltaN}
\Delta_N = (1-Z_{\Gamma_N})\Delta_S.
\end{equation}
Having solved for $\Delta_N$, we now solve for the renormalized superconducting pairing potential in the superconductor. Similar to Eq.~\eqref{eq:sigman}, the self-energy of the superconductor due to electron tunneling from the 2DEG is given by
\begin{align}
\Sigma_S(\omega)& = -|t|^2 \nu(\varepsilon_{F_S}) \left[\frac{\omega\tau_0 + \Delta_N \tau_x }{\sqrt{\Delta_N^2 -\omega^2}} + \zeta_S \tau_z\right].
\end{align}
Substituting this into the strong-coupling form of Green's function of the superconductor, we have
\begin{align}\label{eq:GN}
G_{S}(\boldsymbol{k},\omega) &= \frac{Z_{\Gamma_S}}{\omega - Z_{\Gamma_S}H_N - [Z_{\Gamma_S}\Delta_0 +(1-Z_{\Gamma_S})\Delta_N]\tau_x},
\end{align}
where
\begin{align}
Z_{\Gamma_S}(\omega) = \left(1+\frac{\Gamma_S}{\sqrt{\Delta_N^2 -\omega^2}}\right)^{-1}.
\end{align}
Thus, the renormalized superconducting pairing potential in the superconductor is given by
\begin{equation}\label{eq:DeltaS}
\Delta_S = Z_{\Gamma_S}\Delta_0 + (1-Z_{\Gamma_S})\Delta_N,
\end{equation}
where $\Delta_0$ is the gap of an isolated superconductor. Note that the subscripts $N,S$ in the above equations refer to the quantities in the 2DEG ($N$) and superconductor ($S$), respectively.  The coupled gap equations [Eqs.~\eqref{eq:DeltaN} and~\eqref{eq:DeltaS}] reflect the fact that proximitization is a two-way process. This leads to a pairing gap in a normal material and at the same time it renormalizes the excitation gap in the host superconductor.

\subsection{Relation to the standard effective model}
In the literature, it is rather common to ignore the corrections
in the host superconductor and assume
$\Delta_S = \Delta_0$
but we will see in the full proximitization theory that this
is not generally a good assumption.
Also important is that in the more general situation, 
all pair amplitude parameters vary continuously across the system.

With this simplification, the
above analysis is the basis for the so-called ``effective model" which 
is described as having integrated out the host superconductor. In the effective model, the Hamiltonian of the 2DEG is given by~\cite{Pientka2017Topological,Hell2017Two}
\begin{align}\label{eq:HSM}
\mathcal{H}_{k_x} &=  \left(k_x^2 - \partial_y^2 - \partial_z^2 - \mu \right)\tau_z + \alpha (k_x \sigma_y + i\partial_y\sigma_x)\tau_z\nonumber\\
&\hspace{2 cm} + E_{Z}(y) \sigma_x + \Delta(y)\tau_+ + \Delta^*(y)\tau_-,
 \end{align}
where $\Delta$ is the
proximity-induced pairing potential in the 2DEG which is obtained after integrating out the superconductors. This is given by
\begin{equation}
\Delta(y) = \begin{cases}
\Delta^{\mathrm{prox}} e^{-i\phi/2} & \text{for $-(W_{\mathrm{SC}}+ W/2)< y < -W/2$},\\
0 & \text{for $-W/2 < y < W/2$},\\
\Delta^{\mathrm{prox}} e^{i\phi/2} & \text{for $W/2 < y < W_{\mathrm{SC}}+W/2$},
\end{cases}
\end{equation}
where $\Delta^{\mathrm{prox}}$ is chosen phenomenologically.

\section{Understanding the proximity-induced gap $\Delta^{\mathrm{prox}}$}\label{sec:proximitygap}
\begin{figure}
\includegraphics[width=\linewidth]
{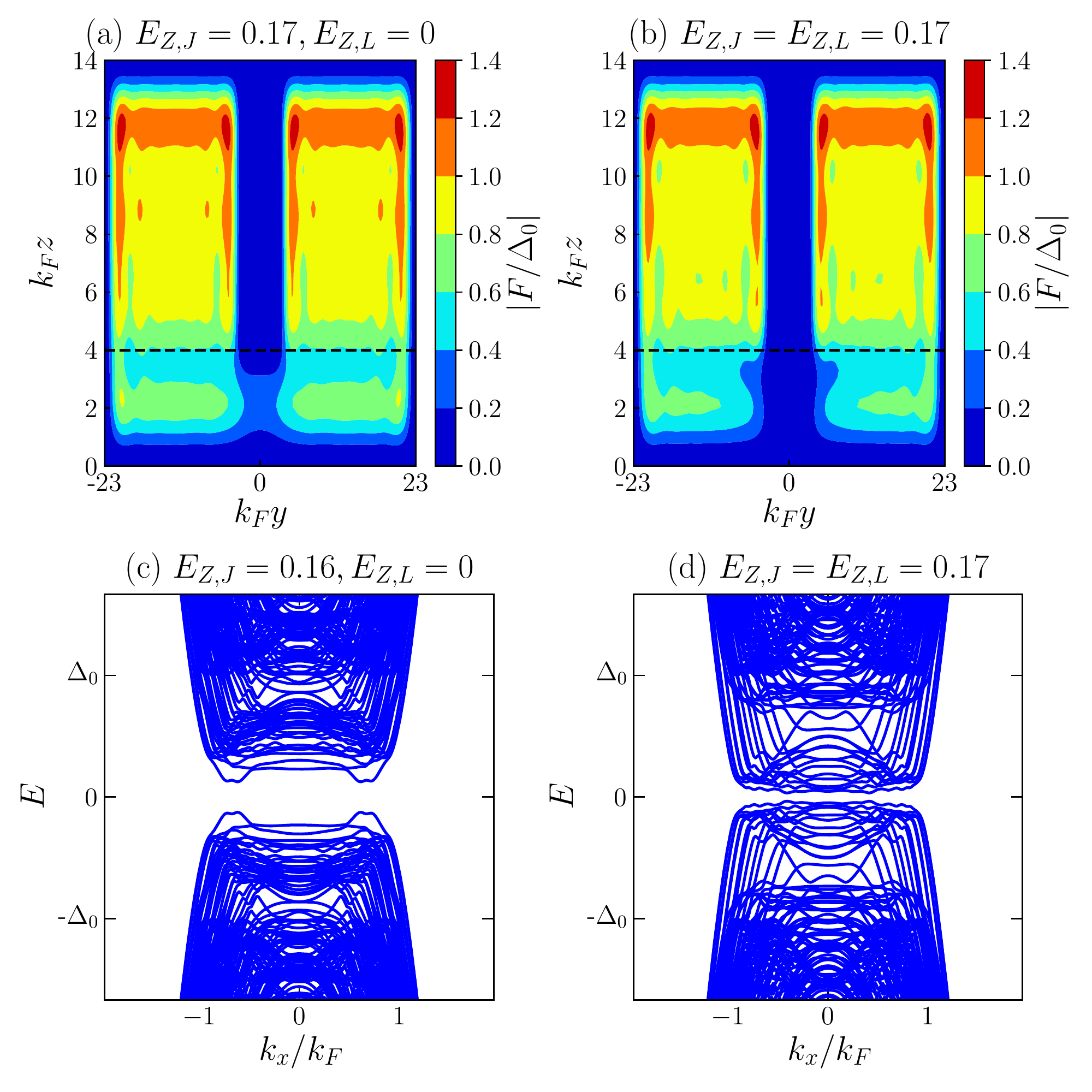}
\caption{Profile of pair amplitudes (top panel) and energy spectra (bottom panel) of the planar Josephson junction for the case: (Left panel) Zeeman is only in the junction ($E_{Z,J} = 0.17$ and $E_{Z,L} = 0$) and (right panel) Zeeman is uniform across the 2DEG ($E_{Z,J} = E_{Z,L} = 0.17$). Note that the presence of the Zeeman field in the 2DEG below the superconductor ($E_{Z,L}$) reduces the induced pair amplitude and proximity gap in the 2DEG [panels (b) and (d)]. The black dashed lines in the top panel denote the boundaries between the superconductors and the 2DEG. The parameters used are $\mu_{\mathrm{S}} = 1$, $\mu_{\mathrm{2DEG}} = 1$, $\alpha = 0.05$, $\Delta_{0}$ = 0.3 [$\xi = v_F/(\pi \Delta_0) = 2.12/k_F$], $\phi =0$, $W_{\mathrm{SC}}$ = 20/$k_{F}$, $W$ = 6/$k_{F}$,  $D_{\mathrm{SC}}$ = 10/$k_{F}$, and $D_{\mathrm{2DEG}}$ = 4/$k_{F}$. }
\label{fig:EZuniform_or_not}
\end{figure}

We turn now to numerical results for
$\Delta^{\mathrm{prox}}$
obtained from our full proximitization studies.
Although we begin with the limit of zero magnetic field, 
it is useful to understand how the magnetic field
affects the separate proximitization processes at each of the two
interfaces between the 2DEG and the host superconductor.
To do this we compare two kinds of Josephson-junction configuration: The first junction has the Zeeman field confined to the channel in the 2DEG between the two superconductors [Figs.~\ref{fig:EZuniform_or_not}(a) and~\ref{fig:EZuniform_or_not}(c)] and the second junction has the
field applied uniformly in the 2DEG substrate [Figs.~\ref{fig:EZuniform_or_not}(b) and~\ref{fig:EZuniform_or_not}(d)], as in experiments.

The upper panels in Fig.~\ref{fig:EZuniform_or_not} present contour plots of the pair amplitudes
and the lower plots show the energy dispersions.
One can see that a magnetic field below the superconductors has very little effect on the parent superconductors but, as expected, it inhibits proximitization and does decrease the pair amplitude and 
energy gap in the 2DEG. Fortunately with the planar Josephson-junction design, we can tune the phase difference towards $\pi$ where the critical field for the transition into the topological phase is smaller such that there is still a substantial gap present when the system is in the topological phase.

In the remainder of this section, we will address how to optimize the proximity
gap $\Delta^{\mathrm{prox}}$ at $ E_Z = \phi = 0$.
By dropping the Zeeman field and junction
phase bias, we are establishing how to select materials as well as geometric parameters.

\subsection{Effects of variable spin-orbit coupling and chemical potential mismatch}\label{sec:soc_gap}
Since SOC plays an important role, it should
be noted that there is no consensus in the literature about
how SOC interacts with proximitization.
It has been argued that larger SOC is beneficial \cite{Potter2011engineering}.
We find here, that in the absence of a magnetic field, the effects of SOC on the proximity-induced gap
are strongly tied to the size of the chemical potential difference between the superconductors
and the 2DEG. 
This can be understood in large part because
of a mismatch in the Fermi momenta of the bands in the superconductors with those of the spin-orbit-coupled 2DEG.

\begin{figure}
\includegraphics[width=\linewidth]
{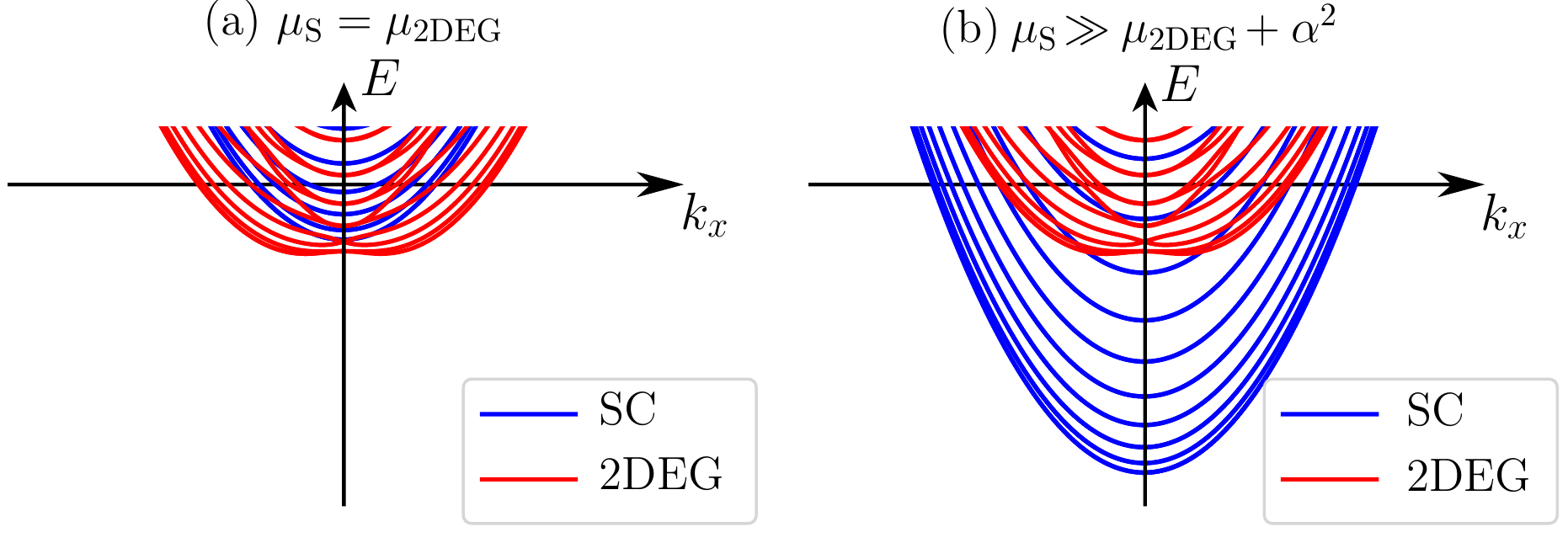}
\caption{Understanding effects of chemical potential mismatch on proximitization. Energy spectra of the normal part of the Hamiltonian of the superconductor (SC) and 2DEG for the case where (a) $\mu_{\mathrm{S}} = \mu_{\mathrm{2DEG}}$ and (b) $\mu_{\mathrm{S}} \gg \mu_{\mathrm{2DEG}} + \alpha^2$. For the case where (a) $\mu_{\mathrm{S}} = \mu_{\mathrm{2DEG}}$, the mismatch between the Fermi momenta of the SC and 2DEG gets larger for increasing SOC strength $\alpha$ while for the case where (b) $\mu_{\mathrm{S}} \gg \mu_{\mathrm{2DEG}} + \alpha^2$, the mismatch between the Fermi momenta of the SC and 2DEG is weakly dependent on the SOC strength $\alpha$. In summary, the dependence of the proximity gap $\Delta^{\mathrm{prox}}$ on $\alpha$ is weaker for the case where the SC chemical potential is much larger than the 2DEG chemical potential.}
\label{fig:mismatchk}
\end{figure}

This mismatch is illustrated in
Fig.~\ref{fig:mismatchk}. Here
the left panel (a) shows the superposed normal-state dispersions for the case where the superconductor and spin-orbit-coupled 2DEG have
the same chemical potential and the right panel (b) is for the case where the chemical potential
in the superconductor is much larger than that in the 2DEG, as is more often the case. 
The principal conclusion from panel (a) is that there are many bands in 2DEG  which have little Fermi momentum overlap (because of the shift due to SOC in the 2DEG) with bands in the superconductors; one can anticipate that this mismatch increases as the SOC becomes larger. This is in contrast to panel (b) where all bands in the 2DEG have their Fermi momenta close to those in the superconductor. Here the deleterious effects of 
SOC on the proximity-induced gap will be less apparent. 

We summarize this by noting that the dependence of the proximitized gap on the SOC strength is weaker for the case where the superconductor chemical potential is larger than the 2DEG chemical potential. This is because
a superconductor with a larger chemical potential has more occupied subbands. As a result, for an incident electron coming from the 2DEG with transverse momentum normal to the $NS$ interface, there is an electron from one of the subbands in the superconductor with momentum which is close to matching the incident momentum of the electron from the 2DEG. 

\begin{figure}[h]
\includegraphics[width=\linewidth]
{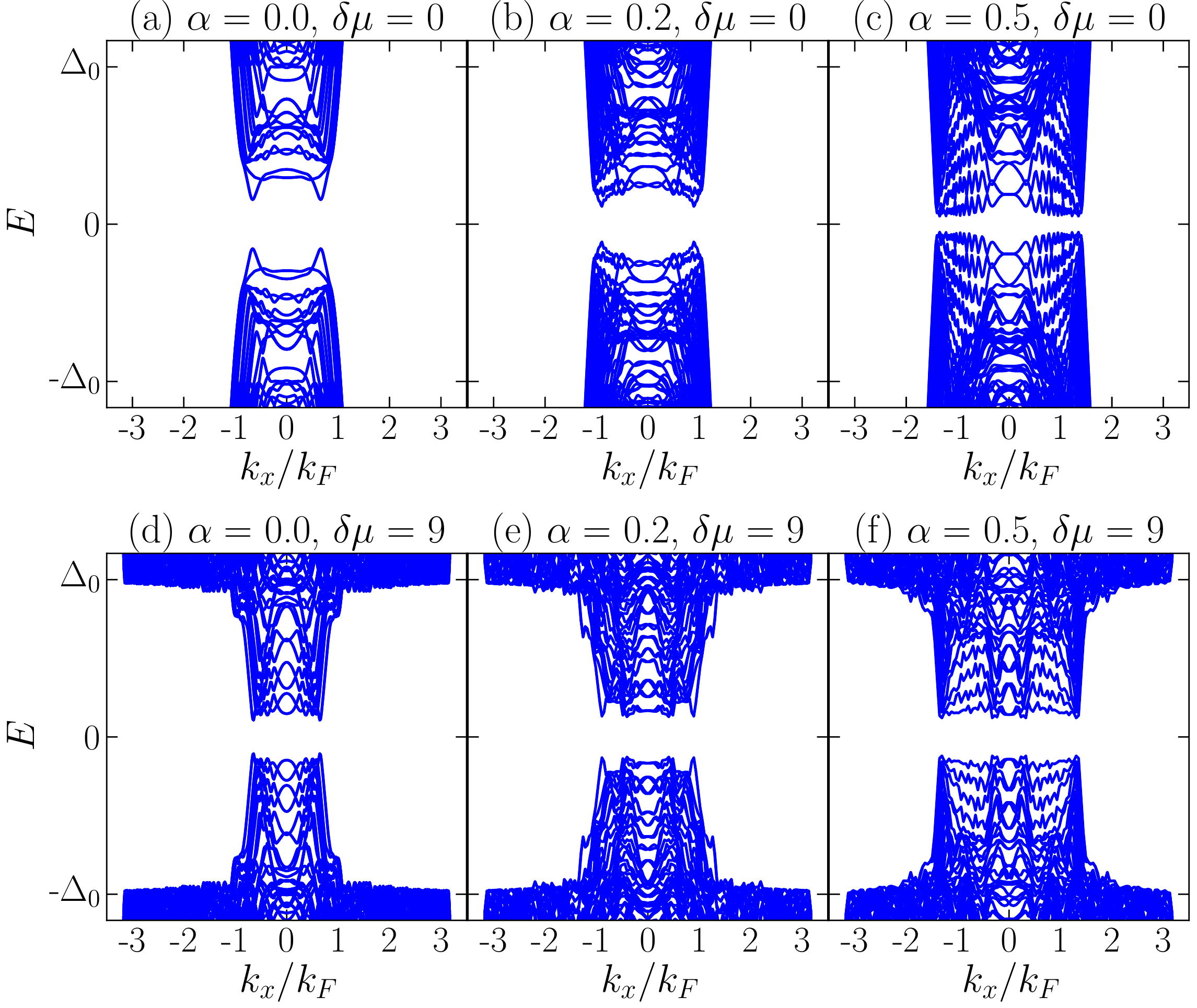}
\caption{Effects of SOC on the spectral gap for the case where there is no magnetic field. Top panel: For small chemical potential mismatch, e.g., $\delta\mu = \mu_{\mathrm{S}} - \mu_{\mathrm{2DEG}}  = 0$, the gap depends strongly on the 2DEG SOC. The gap decreases with increasing SOC strength as shown in panels (a)-(c) because there is a larger mismatch between the Fermi momentum of the superconductor and Rashba spin-orbit-coupled 2DEG as the SOC strength increases. Bottom panel: For large chemical potential mismatch, e.g., $\delta\mu = \mu_{\mathrm{S}} - \mu_{\mathrm{2DEG}}  = 9$, the gap depends weakly on the 2DEG SOC [see panels (d)-(f)] as there are more occupied subbands in superconductors with large $\mu_S$. This implies that for an incident electron coming from one of the bands of the 2DEG, there is a band in the superconductor with a momentum close to the incident momentum. The parameters used are $\mu_{\mathrm{2DEG}} = 1$, $E_{Z,J} =  E_{Z,L} = 0$, $\Delta_0$ = 0.3, $\phi = 0$, $W_{\mathrm{SC}}$ = 20/$k_{F}$, $W$ = 6/$k_{F}$,  $D_{\mathrm{SC}}$ = 10/$k_{F}$, and $D_{\mathrm{2DEG}}$ = 4/$k_{F}$.}
\label{fig:energygapSOC}
\end{figure}

Note that a mismatch in the Fermi velocity of the electron in the superconductor and 2DEG increases the amplitude of the normal reflections while decreasing that of Andreev reflections. Since the superconductivity in the 2DEG is proximity induced via Andreev reflection processes at the interface~\cite{pannetier2000andreev,klapwijk2004proximity}, the mismatch in turn reduces the strength of the proximity-induced gap.

These physical effects are illustrated more directly in Fig.~\ref{fig:energygapSOC}. As shown in the top panel for the case where $\mu_{\mathrm{S}} = \mu_{\mathrm{2DEG}}$, the Fermi momentum mismatch between the superconductors and 2DEG increases as the SOC strength increases in the 2DEG which in turns reduces the proximity gap. The effect of the SOC on the proximity gap is less pronounced for the case where $\mu_{\mathrm{S}} \gg \mu_{\mathrm{2DEG}}$. This is shown in the lower panel of Fig.~\ref{fig:energygapSOC}. 
In summary, for a weaker dependence of the proximity-induced gap on the SOC, the chemical potential of the superconductor has to be much larger than
that of the 2DEG.

But this raises another important issue.
While a substantial
mismatch in chemical potentials helps to negate the SOC effects on the proximitization, there is
a negative side to making the chemical potential mismatch ($\delta\mu = \mu_{\mathrm{S}} - \mu_{\mathrm{2DEG}}$) too large. 
To make this clear, we can compare Figs.~\ref{fig:energygapSOC}(a) and~\ref{fig:energygapSOC}(d) which represent an extreme example of zero SOC in the 2DEG.
Here one can see that the larger the chemical potential difference, the
smaller the effective pairing gap. This is because the chemical potential mismatch increases the Fermi velocity mismatch between the 2DEG and the superconductors resulting in a decrease in the NS interface transparency.~\footnote{The transparency of the interface between the superconductor and 2DEG is determined by the matching of the Fermi velocity between the two materials. While the difference between the Fermi energy of the superconductor and semiconductor can be of several orders of magnitude, the Fermi velocity between the two has a smaller mismatch in magnitude because of a smaller electron effective mass of the semiconductor, which is not accommodated here. We note that the transparency is also affected by microscopic parameters such as the tunnel coupling strength and charge accumulation at the 2DEG-SC interface which are not accounted for in our model.}
We will refer back to these competing effects involving $\delta \mu$ and 
the SOC strength, $\alpha$, in a summary figure (Fig.~\ref{fig:gapthickness}) below, but we here emphasize
the subtle
tradeoffs which must be considered to optimize the
outcome.

\subsection{Effects of variable channel width and variable junction thickness}
\begin{figure}
\includegraphics[width=\linewidth]
{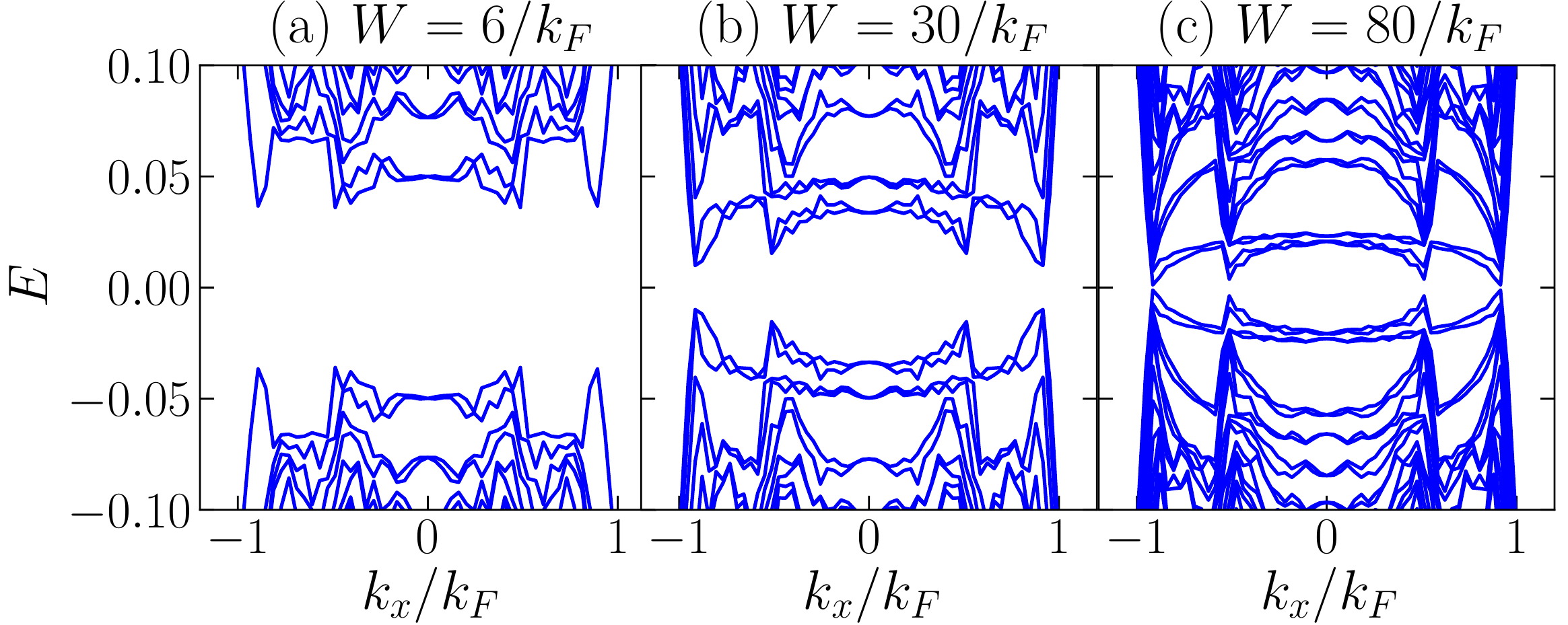}
\caption{Energy spectra of planar Josephson junctions for different junction widths: (a) $W$ = 6/$k_{F}$, (b) $W$ = 30/$k_{F}$, and (c) $W$ = 80/$k_{F}$. The spectral gap decreases with increasing junction width $W$. The parameters used are: $\mu_{\mathrm{2DEG}} = 1$, $\mu_{\mathrm{S}} = 10$, $E_{Z,J} =  E_{Z,L} = 0$, $\Delta_0$ = 0.3 [$\xi = v_F/(\pi \Delta_0) = 2.12/k_F$], $\phi = 0$, $W_{\mathrm{SC}}$ = 20/$k_{F}$,  $D_{\mathrm{SC}}$ = 10/$k_{F}$, and $D_{\mathrm{2DEG}}$ = 4/$k_{F}$.}
\label{fig:diffwidth}
\end{figure}

Figure~\ref{fig:diffwidth} illustrates a striking effect of increasing the width
of the quasi-1D channel of the junction in the 2DEG.  The pairing gap is greatly suppressed as the
channel becomes wider. This is relatively easy to understand, as proximitization strength (arising from the leaking of Cooper pairs from the superconductors to 2DEG)
decays with increasing distance from the superconductors which results in a smaller superconducting gap for a wider junction between the two superconductors. We illustrate this
case in part because this wide channel situation is more favorable for observing
the FFLO phase discussed in Sec.~\ref{sec:FFLO}.

\begin{figure}
\includegraphics[width=\linewidth]
{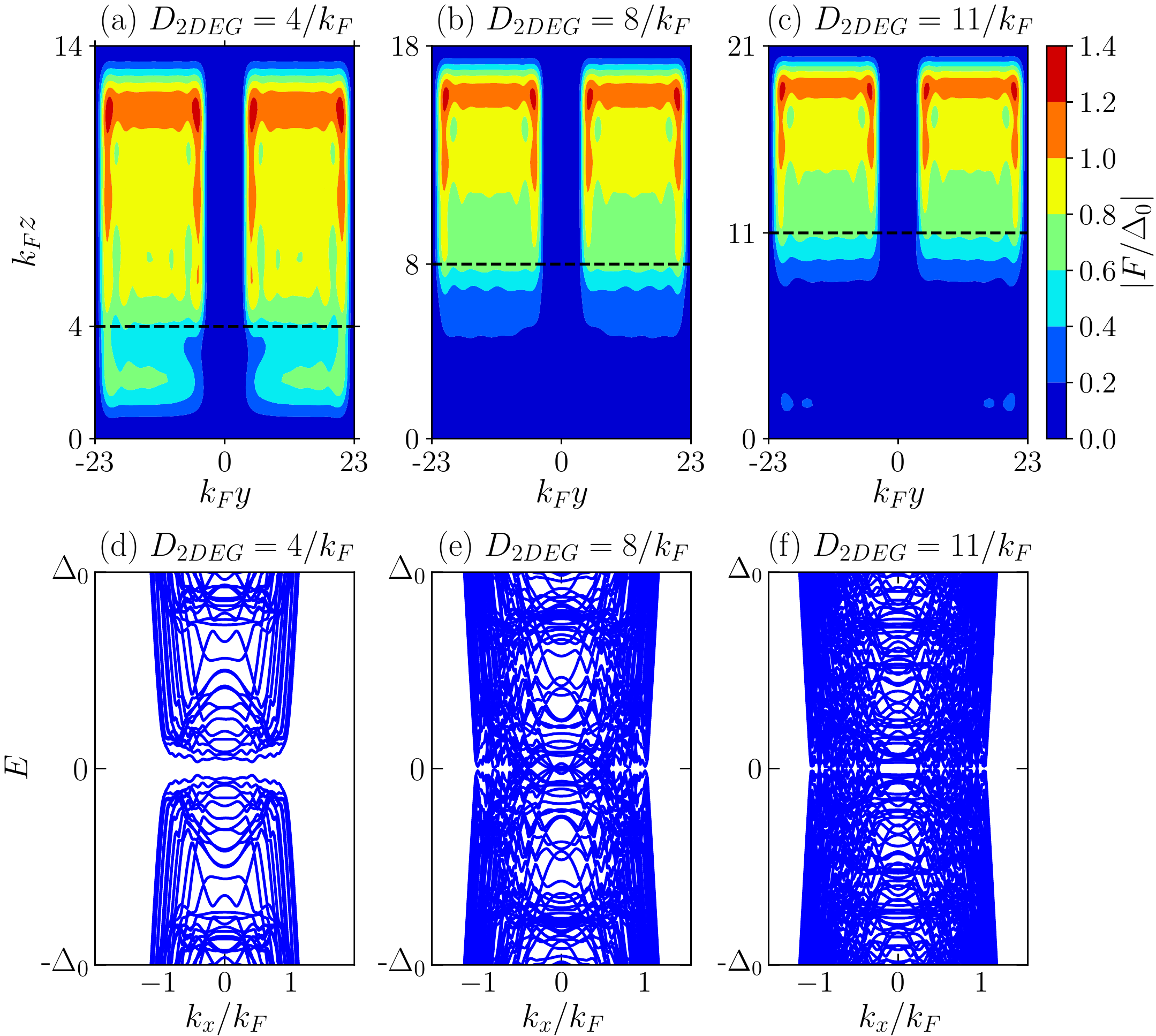}
\caption{Thickness effects. Profile of pair amplitude (top panel) and energy spectra (bottom panel) of planar Josephson junctions for zero Zeeman field and different thickness of 2DEG: $D_{\mathrm{2DEG}} = 4/k_F$ (left panel), $D_{\mathrm{2DEG}} = 8/k_F$ (middle panel)   and $D_{\mathrm{2DEG}} = 11/k_F$ (right panel). Note that the thicker the 2DEG is, the smaller is the induced superconducting gap in the 2DEG. The black dashed lines in the top panel denote the boundaries between the superconductors and the 2DEG. The parameters used are $\mu_{\mathrm{S}} = 1$, $\mu_{\mathrm{2DEG}} = 1$, $\alpha = 0.05$, $E_{Z,J} = E_{Z,L} = 0$, $\Delta_{0}$ = 0.3 [$\xi = v_F/(\pi \Delta_0) = 2.12/k_F$], $\phi =0$, $W_{\mathrm{SC}}$ = 20/$k_{F}$, $W$ = 6/$k_{F}$,  and $D_{\mathrm{SC}}$ = 10/$k_{F}$. }
\label{fig:thickness}
\end{figure}

\begin{figure*}
\includegraphics[width=5.5in,clip]
{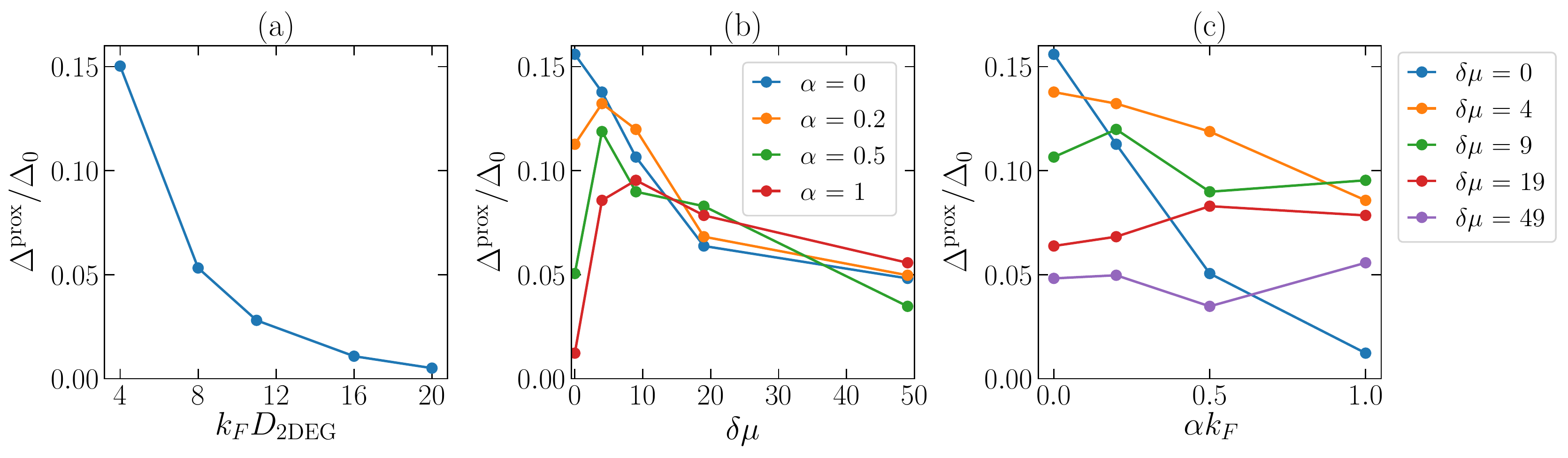}
\caption{Summary figure showing how 
$\Delta^{\mathrm{prox}}$ depends on 2DEG thickness, $\delta \mu$, and SOC. 
(a) Induced gap $\Delta^{\mathrm{prox}}/\Delta_0$ as a function of the 2DEG thickness $D_{\mathrm{2DEG}}$ for SOC strength $\alpha = 0.05$ and $\mu_{\mathrm{S}} = \mu_{\mathrm{2DEG}} = 1$. The induced gap decreases with increasing 2DEG thickness. (b) Induced gap $\Delta^{\mathrm{prox}}/\Delta_0$ as a function of chemical potential difference ($\delta\mu = \mu_{\mathrm{S}} - \mu_{\mathrm{2DEG}}$) calculated for $\mu_{\mathrm{2DEG}} = 1$, $D_{\mathrm{2DEG}} = 4/k_F$ and several values of SOC strength $\alpha$. For small $\alpha$, the induced gap decreases with increasing $\delta \mu$. For large $\alpha$, the induced gap has nonmonotonic dependences on $\delta \mu$ where it first increases with increasing $\delta \mu$, rises to a maximum, and after reaching the maximum it decreases with increasing $\delta \mu$. (c) Induced gap $\Delta^{\mathrm{prox}}/\Delta_0$ as a function of SOC strength $\alpha$ for $D_{\mathrm{2DEG}} = 4/k_F$ and different values of $\delta \mu$. For small $\delta \mu$, the induced gap depends strongly on $\alpha$ where it decreases with increasing $\alpha$. For the case where $\mu_{\mathrm{S}}$ is much bigger than $\mu_{\mathrm{2DEG}}$, the induced gap depends weakly on $\alpha$. The parameters used for the above plots are $W_{\mathrm{SC}} = 20/k_{\mathrm{F}}$, $W = 6/k_{\mathrm{F}}$, $D_{\mathrm{SC}} = 10/k_{\mathrm{F}}$, $E_{Z,J} = E_{Z,L} = 0$, $\Delta_0 =0.3$, and $\mu_{\mathrm{2DEG}} = 1$.}
\label{fig:gapthickness}
\end{figure*}

Figure~\ref{fig:thickness} addresses the effect of varying the thickness of the 2DEG on the proximity gap, illustrating another effect associated with geometry.
Shown here are plots of
the pair amplitude (upper panel) and energy spectra (lower panel) of the Josephson junction. It can be seen from the plots that the pair amplitude and spectral gap decrease with increasing thickness of the 2DEG. There are contrary suggestions in
the literature~\cite{Lutchyn2011Search,Stanescu2011Majorana} that these thicker substrates could be
favorable as they allow ``multichannel participation". As shown here, though,
thicker junctions lead to smaller proximity gaps since they require that the superconducting correlations
extend over a greater distance deeper into the 2DEG. 
We, thus, conclude that as
Majorana zero modes are protected by large proximity-induced gaps, thinner 2DEGs 
are more favorable to be used as platforms for topological quantum computation.  

Figure~\ref{fig:gapthickness} presents a summary of how $\Delta^{\mathrm{prox}}$ is affected by geometry and materials parameters. This figure shows how increasing (a) the thickness, (b) the chemical potential difference and (c) the SOC strength affect the proximity gap (at zero Zeeman field and zero phase difference). Clearly making both the thickness and the channel width larger has deleterious effects. However, as shown in Figs.~\ref{fig:gapthickness}(b) and~\ref{fig:gapthickness}(c), the effects of SOC are strongly connected to the magnitude of the chemical potential difference ($\delta \mu = \mu_{\mathrm{S}} - \mu_{\mathrm{2DEG}}$)~\footnote{Note that the realistic picture of chemical potential mismatch can be quite complicated as it involves band bending and charge screening at the NS interface (which are not taken into account in our model). Taking the band bending into account implies that the larger the chemical potential mismatch between the 2DEG and the superconductor, the larger Rashba SOC field (due to larger
electrical field at the interface).}. As shown in Fig.~\ref{fig:gapthickness}(b), when there is any finite SOC, there is a notable nonmonotonicity in plots of $\Delta^{\mathrm{prox}}$ versus $\delta \mu$. The initial rise in $\Delta^{\mathrm{prox}}$ with $\delta\mu$ for a fixed $\alpha$ is due to the matching of the band structure of the superconductor with the Rashba-derived band structure in the 2DEG. However, once the chemical potential difference is sufficiently large, as might be expected, increasing it further has a negative effect on the proximity gap due to the mismatch in the Fermi momenta between the superconductors and 2DEG, as illustrated in Fig.~\ref{fig:mismatchk}. There seems to be a ``sweet spot" around $\delta \mu \approx 10$ which is substantially below the more realistic physical regime (where $\delta \mu$ might approach $100$ or larger). Figure~\ref{fig:gapthickness}(c) shows that the effects of SOC on the proximity gap $\Delta^{\mathrm{prox}}$ becomes weaker as $\delta \mu$ increases, as discussed in Sec.~\ref{sec:soc_gap}. Overall this figure should help guide materials parameters and geometries~\footnote{We note that our results are obtained using parabolic dispersions for the semiconductor and superconductor. In general, the semiconductor band structure is well approximated by a parabolic band dispersion since the chemical potential of a semiconductor is close to the band bottom. Our choice of the band structure for superconductor will not affect our results as the effect of the band structure of the superconductor is independent of the effect of the above three parameters (thickness, spin-orbit coupling, and chemical potential mismatch) on the proximity gap.}.

\section{Symmetry class}\label{sec:symmetry}
It is useful to look at the underlying symmetries which dictate the nature of the topological phases. The above BdG Hamiltonian~[Eq.~\eqref{eq:BdG}] for the planar Josephson junction commutes with the particle-hole symmetry operator $\mathcal{P} = \sigma_y \tau_y \mathcal{K}$ where $\mathcal{K}$ is the complex conjugation. For zero Zeeman field $E_Z = 0$ and a superconducting phase bias $\phi = 0$ or $\phi =\pi$, the Hamiltonian belongs to the symmetry class DIII in the tenfold classification~\cite{kitaev2009periodic,ryu2010topological,Alexander1997Nonstandard} as it also commutes with the time-reversal symmetry operator $\mathcal{T} = -i\sigma_y \mathcal{K}$ (where $\mathcal{T}^2 = -\mathds{1}$). Moreover, the system also has a mirror symmetry along the $x$-$z$ plane with the mirror operator given by $\mathcal{M}_y = -\sigma_y \times (y \rightarrow -y)$.

The $\mathcal{T}$ and $\mathcal{M}_y$ symmetries are broken when an in-plane Zeeman field is applied along the junction ($x$ direction) or for a phase bias other than $\phi = 0$ or $\phi=\pi$. The Hamiltonian, however, remains invariant under an antiunitary ``effective" time-reversal
operator $\widetilde{\mathcal{T}}$ which is the product of the $\mathcal{T}$ and $\mathcal{M}_y$ operators, i.e.,  $\widetilde{\mathcal{T}} = \mathcal{M}_y \mathcal{T} = i \mathcal{K} \times (y \rightarrow -y)$ where $\mathcal{T}^2 = \mathds{1}$. Thus the 
system has the BDI symmetry~\cite{Pientka2017Topological,Hell2017Two}. Moreover, since the Hamiltonian possesses $\widetilde{\mathcal{T}}$
and $\mathcal{P}$ symmetries, it also has a chiral symmetry, where the Hamiltonian anticommutes with the chirality operator $\mathcal{C} = -i \mathcal{P}\widetilde{\mathcal{T}} = \mathcal{M}_y \tau_y$. When the $\widetilde{\mathcal{T}}$ symmetry is broken, the symmetry class is reduced from class BDI to class D. In this case, an even number of Majorana zero modes at the same end of the junction couples to each other and splits into finite-energy mode leaving either zero or one Majorana mode at each end of the junction. This BDI symmetry can be broken by disorder~\cite{Haim2018the}, applying a transverse Zeeman field perpendicular to the junction (along the $y$ direction)~\cite{setiawan2019topological} or having left and right superconductors with different widths or pairing potentials~\cite{Pientka2017Topological,Hell2017Two,setiawan2019topological}.

The symmetry class BDI is characterized by a $\mathbb{Z}$ topological invariant $Q_{\mathbb{Z}}$ where $|Q_{\mathbb{Z}}|$ denotes the number of Majorana zero modes at each end of the junctions. On the other hand, the symmetry class D is characterized by a $\mathbb{Z}_2$ topological invariant $Q_{\mathbb{Z}_2}$ which denotes the parity of the $Q_{\mathbb{Z}}$ invariant.

\section{Topological phase diagram and transition}\label{sec:topological}

We obtain the phase diagram of the system by calculating the topological invariant
following Ref.~\cite{Tewari2012Topological}.
The numerical computation is considerably more complicated in the presence of our full treatment of
proximitization. To do so,
we first diagonalize the chiral operator $\mathcal{C}$ with $\mathds{1}$ and $-\mathds{1}$ in the upper-left and lower-right block, respectively. Since $\{\mathcal{C},\mathcal{H}\} = 0$, in this basis where the $\mathcal{C}$ is block diagonal, the BdG Hamiltonian $\mathcal{H}_{k_x}$ is off diagonal, i.e.,
\begin{subequations}
\begin{align}
U\mathcal{C}U^\dagger &= \left(\begin{matrix} \mathds{1} & 0 \\ 0 & -\mathds{1} \end{matrix}\right), \\ 
U\mathcal{H}_{k_x}U^\dagger &= \left(\begin{matrix} 0 & A(k_x) \\ A^T(-k_x) & 0 \end{matrix}\right).
\end{align}
\end{subequations}
We can calculate the $\mathbb{Z}$ topological invariant ($Q_{\mathbb{Z}}$) from the winding of the phase $\theta(k_x)$ of the determinant of the off-diagonal part $A(k_x)$ where $e^{i\theta(k_x)} = \mathrm{det}A(k_x)/|\mathrm{det}A(k_x)|$. The $\mathbb{Z}$ topological invariant is given by 
\begin{equation}
Q_{\mathbb{Z}} = \int_0^{\infty} \frac{d k_x}{\pi} \frac{d\theta(k_x)}{dk_x},
\end {equation}
and the $\mathbb{Z}_2$ topological invariant (the parity of $Q_\mathbb{Z}$) is given by
\begin{equation}\label{eq:QZ2}
Q_{\mathbb{Z}_2} = (-1)^{Q_{\mathbb{Z}}}.
\end{equation}
It is shown in Ref.~\cite{Tewari2012Topological} that Eq.~\eqref{eq:QZ2} is simply the $\mathbb{Z}_2$ Pfaffian invariant of 1D systems~\cite{Kitaev2001Unpaired}, i.e.,
\begin{equation}
Q_{\mathbb{Z}_2} = \mathrm{sgn}\frac{\mathrm{Pf}[(\mathcal{H}_{k_x \rightarrow \infty})\sigma_y\tau_y]}{\mathrm{Pf}
[(\mathcal{H}_{k_x =0})\sigma_y\tau_y]}.
\end{equation}

\begin{figure}
\includegraphics[width=\linewidth]
{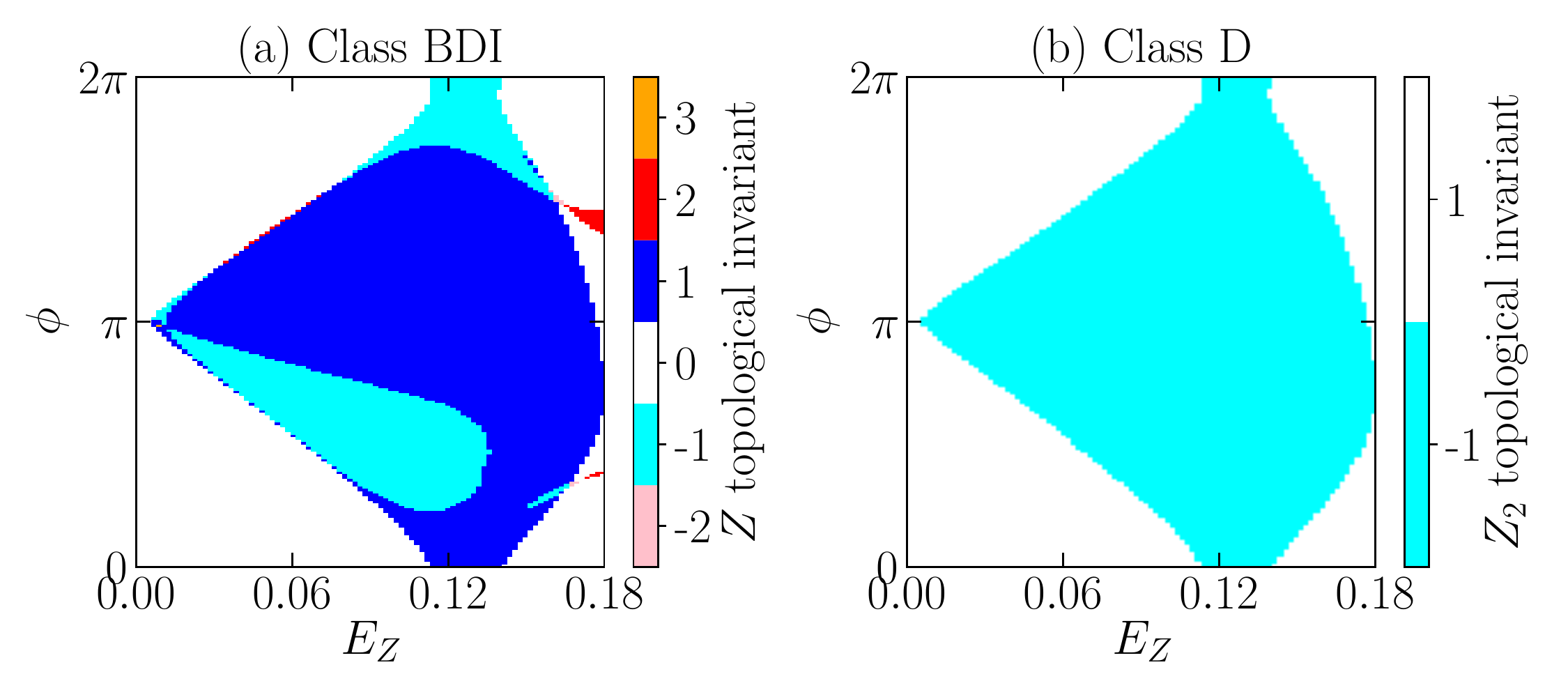}
\caption{Topological phase diagrams for a proximitized Josephson junction with vanishing chemical potential mismatch $\delta \mu = 0$. (a) Class BDI and (b) Class D phase diagram as functions of $E_{Z}$ and $\phi$. Each region is labeled by different $\mathbb{Z}$ topological invariants in the BDI phase diagram. The $\mathbb{Z}_2$ invariant gives the parity of the $\mathbb{Z}$ index. The topological invariant $Q_{\mathbb{Z}_2} = -1$ and $Q_{\mathbb{Z}_2}$ = 1 corresponds to the odd and even $\mathbb{Z}$ indices which in turn indicates the topological and trivial phases of class D. The parameters used are $\mu_{\mathrm{S}} = \mu_{\mathrm{2DEG}} = 1$, $\alpha = 0.05$, $\Delta_0$ = 0.3, $W_{\mathrm{SC}}$ = 20/$k_{F}$, $W$ = 6/$k_{F}$,  $D_{\mathrm{SC}}$ = 10/$k_{F}$, and $D_{\mathrm{2DEG}}$ = 4/$k_{F}$.}
\label{fig:phasediagram}
\end{figure}

\begin{figure}
\includegraphics[width=\linewidth]
{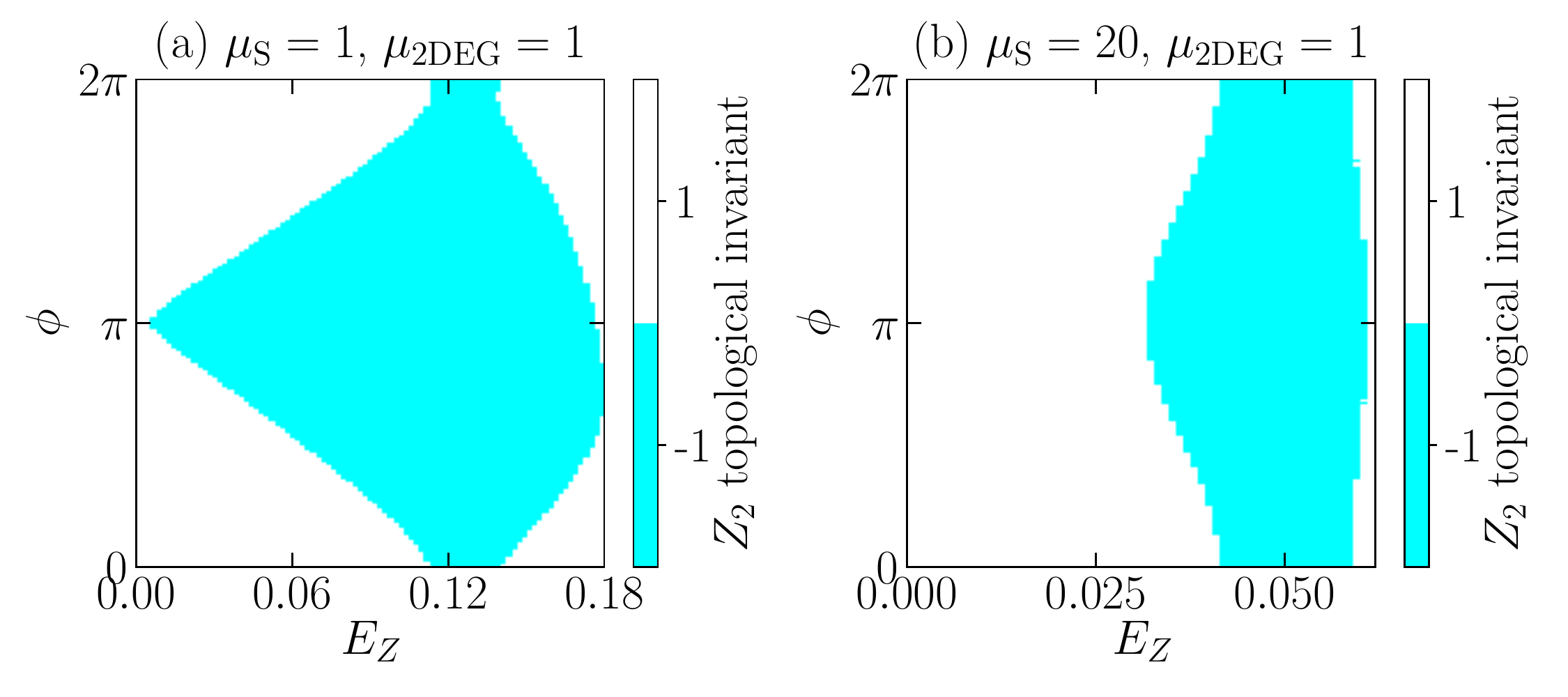}
\caption{Comparison of topological phase diagrams (a) without ($\mu_{\mathrm{S}} = 1$, $\mu_{\mathrm{2DEG}} = 1$) and (b) with  chemical potential mismatch ($\mu_{\mathrm{S}} = 20$, $\mu_{\mathrm{2DEG}} = 1$).
Class D phase diagrams for two different values of chemical potential differences between the superconductors and the 2DEG: (a) $\mu_{\mathrm{S}} = 1$ and $\mu_{\mathrm{2DEG}} = 1$ and (b) $\mu_{\mathrm{S}} = 20$ and $\mu_{\mathrm{2DEG}} = 1$. The $\mathbb{Z}_2$ invariant $Q_{\mathbb{Z}_2}$ = -1 and $Q_{\mathbb{Z}_2}$ = 1 indicate the topological and trivial phases of class D. The chemical potential of the 2DEG is renormalized by the chemical potential of the superconductors resulting in a difference between the effective chemical potential of the 2DEG below the superconductor and that of the 2DEG in the junction. This difference increases as the mismatch between the superconductor and 2DEG chemical potential becomes larger which in turn increases the amplitude of normal reflections in the 2DEG. As a result, for a larger chemical potential mismatch, the phase diagram becomes more stripe like (less dependent on $\phi$) and the critical Zeeman field for $\phi = \pi$ shifts to a larger value. The parameters used are $\mu_{\mathrm{2DEG}} = 1$, $\alpha = 0.05$, $\Delta_0$ = 0.3, $W_{\mathrm{SC}}$ = 20/$k_{F}$, $W$ = 6/$k_{F}$,  $D_{\mathrm{SC}}$ = 10/$k_{F}$, and $D_{\mathrm{2DEG}}$ = 4/$k_{F}$.}
\label{fig:phasediagram2}
\end{figure}

Figures ~\ref{fig:phasediagram} 
and ~\ref{fig:phasediagram2} present the phase diagrams of the planar Josephson junction obtained from the full proximity 
calculations. 
These phase diagrams emphasize the novel feature of the Josephson-junction architecture which enables
the topological phase to be tuned either by changing the phase bias or the Zeeman field.

Figure~\ref{fig:phasediagram} shows the class BDI and class D phase diagrams for the same junction. Each phase in the BDI phase diagram [Fig.~\ref{fig:phasediagram}(a)] is labeled by a different $\mathbb{Z}$ topological invariant ($Q_{\mathbb{Z}}$) where $|Q_\mathbb{Z}|$ denotes the number of Majorana zero modes located at each end of the junction. As can be seen from Fig.~\ref{fig:phasediagram}(a), the $\mathbb{Z} = 1$ topological region occupies most of the phase diagram as it occurs in a wide range of parameters. The topological transition between each of the BDI phases is indicated by a gap closing at $k_x = k_F$. The class D phase diagram~[Fig.~\ref{fig:phasediagram}(b)], on the other hand, shows the parity of the $\mathbb{Z}$ topological invariant [Eq.~\eqref{eq:QZ2}] where $Q_{\mathbb{Z}_2} = 1$ and $Q_{\mathbb{Z}_2} = -1$ correspond to the trivial and topological phases of class D, respectively. The topological transition between the $Q_{\mathbb{Z}_2} = 1$ and $Q_{\mathbb{Z}_2} = -1$ regions is reflected in a gap closing at $k_x = 0$~\cite{Kitaev2001Unpaired} (see Sec.~\ref{sec:energydispersion}). 

The bulk-boundary correspondence implies that the change in the topological index from $Q_{\mathbb{Z}_2} = 1$ to $Q_{\mathbb{Z}_2} = -1$ which is accompanied by a bulk gap closing at $k_x =0$ corresponds to the appearance of a Majorana zero mode at the end of a finite-length junction. Importantly, the edge states that appear in a finite-length junction of our model are Majorana zero modes and \textit{not} Andreev bound states. Clearly, Andreev bound states do not involve a change of topological index (from trivial to topological) and are also not accompanied by bulk gap closings.

Figure~\ref{fig:phasediagram2} shows the effect of chemical potential mismatch ($\delta \mu = \mu_{\mathrm{SC}} - \mu_{\mathrm{2DEG}}$) on the phase dependence of the class D phase diagram. For an ideal or ``transparent" Josephson junction, the phase diagram has a diamond shape where the critical Zeeman field at which the topological phase transition happens is considerably smaller for $\phi = \pi$ than for $\phi = 0$ [see Fig.~\ref{fig:phasediagram2}(a)].  
We observe that, with a larger value for $\delta \mu$, the phase diagram appears to be more stripelike 
as in Fig.~\ref{fig:phasediagram2}(b). 
Here, the dependence of the phase diagram on the superconducting phase difference $\phi$ becomes weaker and the critical Zeeman field for $\phi = \pi$ shifts to a larger value.

We understand this stripelike phase diagram as
deriving from an increasing mismatch between the chemical potential of the superconductor and
the 2DEG. This, in turn, should be viewed as leading to an increase in the strength of the normal reflections in the 2DEG.
Due to the proximity to the superconductor, the chemical potential of the 2DEG directly in contact with the superconductor will be renormalized by that of the superconductor. As a result, there is a difference between the effective chemical potential of the 2DEG directly below the superconductor with the effective chemical potential of the 2DEG in the junction. This effectively creates a potential barrier for the electrons which in turn increases the strength of normal reflections.

We conclude this section by noting that under ideal circumstances (i.e., for transparent junctions with small $\delta \mu$),
the critical
Zeeman field needed to tune the system to topological phases can be greatly reduced for a phase bias $\phi = \pi$.  
One can infer from 
Fig.~\ref{fig:phasediagram2}(b), that when $\delta \mu$ assumes a substantial
(and physically reasonable) value, this gain in reduction of the critical Zeeman field (by tuning the phase $\phi$ to be near $\pi$) is mostly
lost~\footnote{It can be noted that we use the same effective electron mass in the superconductor and 2DEG
in addressing proximitization within this paper, and, thus, we may be
overestimating the reduction in the phase dependence of the class D phase diagram and the proximity-induced gap due to the chemical potential mismatch $\delta \mu$.}. We note that similar to the effect of $\delta \mu$, decreasing the width of the superconducting leads also makes the phase diagram becomes less dependent on the phase bias due to the enhancement of multiple normal reflections at the interface between the superconductors and the vacuum~\cite{setiawan2019topological}.

\subsection{Energy dispersion across the topological phase transition}\label{sec:energydispersion}
The topological phase transition of class D is associated with a gap closing at $k_x = 0$~\cite{Kitaev2001Unpaired}. As can be seen from the phase diagram [Fig.~\ref{fig:phasediagram2}(a)], for a transparent junction the critical Zeeman field at which the transition happens is much smaller when the superconducting phase difference $\phi$ is near $\pi$. As a complement to this phase diagram, we address the energy spectrum of the system as a function of $k_x$ across the phase transition. 

\begin{figure}
\includegraphics[width = \linewidth]
{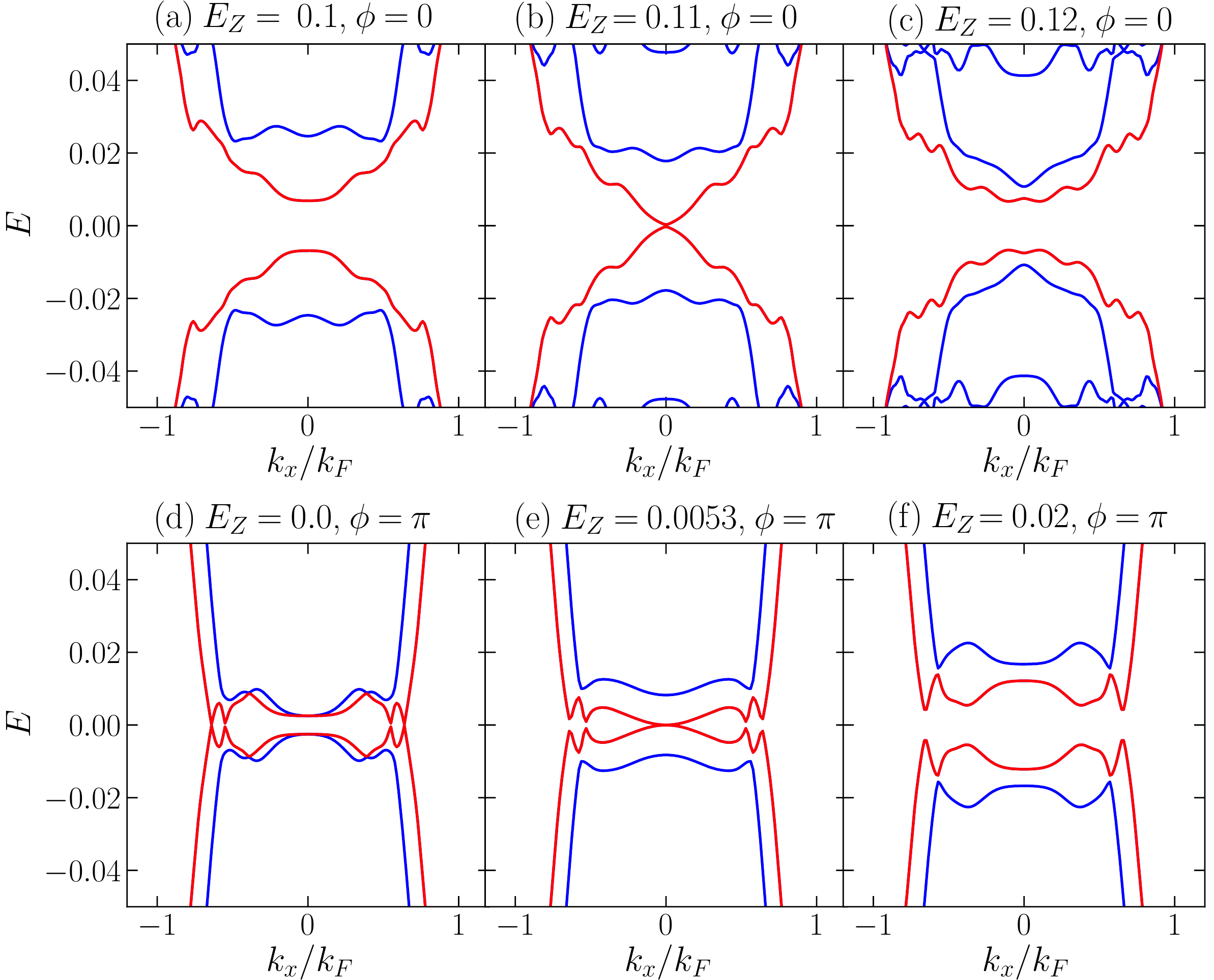}
\caption{Evolution of the energy spectrum of a planar Josephson junction across the topological phase transition for $\phi = 0$ (upper panel) and $\phi = \pi$ (lower panel). The topological transition is characterized by a gap closing at $k_x = 0$ where the critical field at which the topological transition occurs is the smallest at $\phi = \pi$. The critical fields for $\phi = 0$ and $\phi = \pi$ are $E_{Z} = 0.11$ [panel (b)] and $E_{Z} = 0.0053$ [panel (e)], respectively. Energy spectra shown correspond to the phase diagram of Fig.~\ref{fig:phasediagram}. The gap closes and reopens at $k_x = 0$ as the Zeeman field $E_Z$ is respectively tuned towards and away from the critical field.  (a),(d) The system is in the trivial phase; (b),(e) the system undergoes a topological phase transition with a gap closing at $k_x= 0$; (c),(f) the system is in the topological phase.  Shown here are only a few low-energy states close to zero energy where the energy levels closest to zero energy are shown by red lines. Here, we take the Zeeman field to be uniform ($E_{Z,J} = E_{Z,L} = E_Z$) in the 2DEG.  The parameters used are $\mu_{\mathrm{S}} = 1$, $\mu_{\mathrm{2DEG}} = 1$, $\Delta_0$ = 0.3, $\alpha = 0.05$, $W_{\mathrm{SC}}$ = 20/$k_{F}$, $W$ = 6/$k_{F}$,  $D_{\mathrm{SC}}$ = 10/$k_{F}$, and $D_{\mathrm{2DEG}}$ = 4/$k_{F}$. }
\label{fig:Evol_Ez}
\end{figure}

Figure~\ref{fig:Evol_Ez} shows the evolution of the energy spectrum of a planar Josephson junction as the Zeeman field is tuned across the topological phase transition for two different values of superconducting phase differences: $\phi = 0$ (upper panel) and $\phi = \pi$ (lower panel). At a particular value of critical field $E_Z$, the gap at $k_x =0$ closes [panels (b) and (e)] which reflects the transition between trivial and topological phases. The critical Zeeman field is reduced as $\phi \rightarrow \pi$. 

We summarize this section by noting that
despite the more indirect form of proximitization associated with this Josephson-junction
architecture, 
as compared with the nanowires of
Fig.~\ref{fig:fig1}(b),
we have presented strong evidence that proximitized topological phases exist.
This topological superconductivity occurs even when there are no direct attractive interactions
in the 2DEG channel. Nevertheless, in this Josephson-junction
configuration the proximity coupling guarantees that there is a finite pair amplitude within the channel.

\section{Proximity-induced FFLO phase}\label{sec:FFLO}
An exotic superconducting state, characterized by nonzero center-of-mass momentum of Cooper pairs and spatially varying order parameter, may occur for certain materials in the presence of both in-plane magnetic field and superconductivity.
Interestingly, the planar junctions discussed here
are associated with this exotic form of superconductivity, referred to as the Fulde-Ferrell-Larkin-Ovchinikov (FFLO) phase~\cite{Fulde1964Superconductivity,Larkin1965Nonuniform}. 
Indeed, it is hard to find examples where this elusive phase, deriving from magnetic field effects,
has been observed ~\cite{chen2018finite} which do \textit{not} originate from proximity coupling.
Experiments based on this Josephson-junction architecture~\cite{hart2017controlled} report that the FFLO phase
appears to be confined within the 1D channel of the junction.
One might have expected it to be present in some form throughout the 2DEG
since magnetic fields and proximity coupling are present outside the channel as well. Due to the close proximity to the parent superconductor the induced gap there, however, is stronger and is not energetically favorable to oscillate in response to an applied in-plane Zeeman field.  The channel in the junction, on the other hand, is well away from the host superconductors and thus has greatly weakened pair amplitude with superconducting phases which are freer to oscillate.

\begin{figure}
\includegraphics[width=\linewidth]
{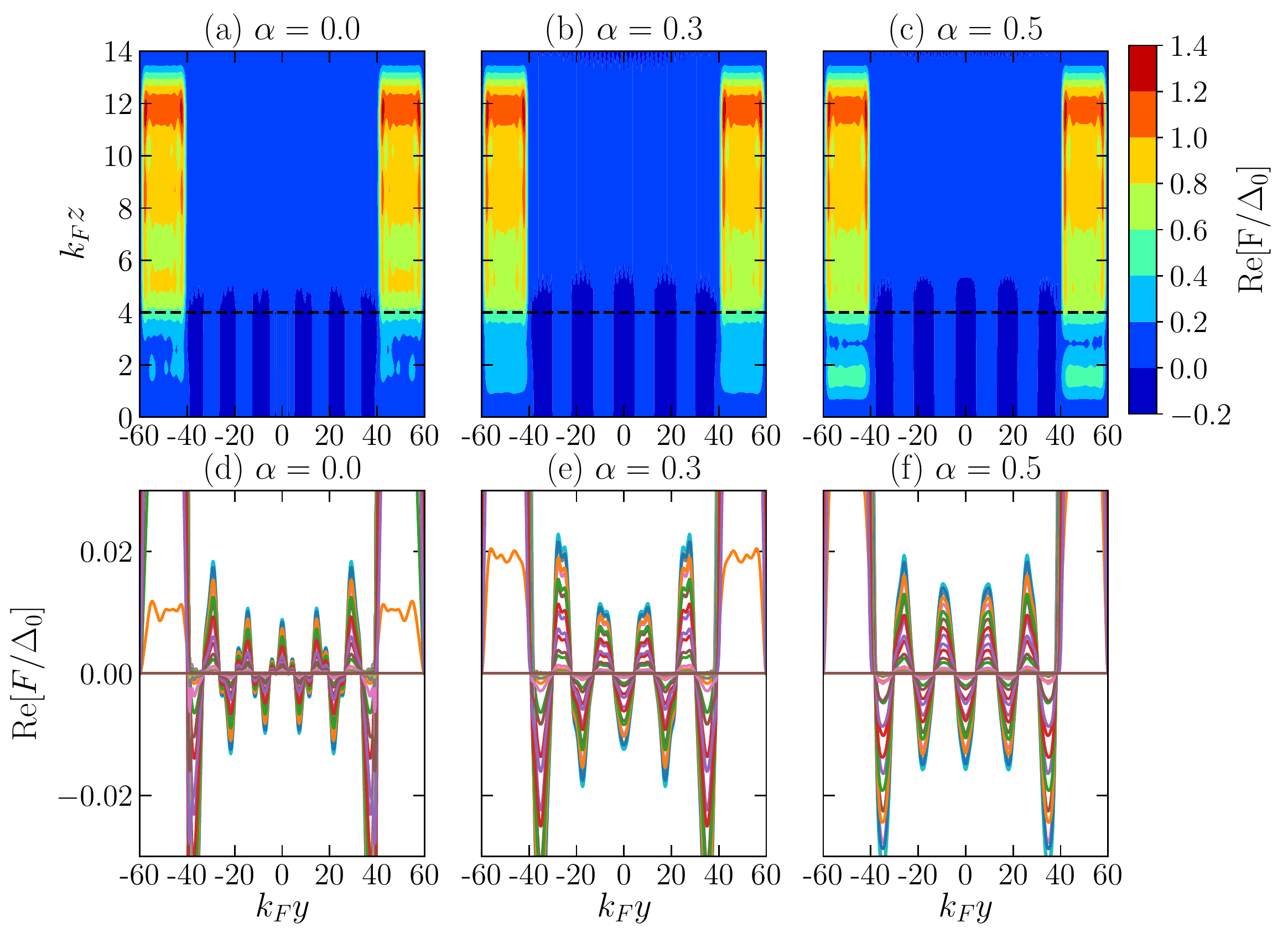}
\caption{Evidence for FFLO pairing. Profile of the real part of the pair amplitude of planar Josephson junctions with 2D Rashba SOC of different strength: $\alpha = 0$ (left panel), $\alpha = 0.3$ (middle panel), and $\alpha = 0.5$ (right panel). Upper panel shows the color plots of the real part of the pair amplitude Re[$F/\Delta_0$],  and the lower panel is  the linecuts of the pair amplitude along the $y$ direction at different values of $z$. As shown in the bottom panel, the pair amplitude oscillates along the $y$ direction with a characteristic oscillation length $\lambda = E_Z/(2v_F)$ that decreases with increasing $\alpha$ (as $v_F$ increases with increasing $\alpha$). The black dashed lines in the upper panel denote the boundary between the superconductors and the 2DEG. The parameters used are $\mu_{\mathrm{S}} = 1$, $\mu_{\mathrm{2DEG}} = 1$, $E_{Z,J} = E_{Z,L} = 0.3$, $\Delta_{0}$ = 0.3, $\phi =0$, $W_{\mathrm{SC}}$ = 20/$k_{F}$, $W$ = 80/$k_{F}$,  $D_{\mathrm{SC}}$ = 10/$k_{F}$, and $D_{\mathrm{2DEG}}$ = 4/$k_{F}$.}
\label{fig:gap2DSOC}
\end{figure}

The upper panel of Fig.~\ref{fig:gap2DSOC} presents a contour plot of the pair amplitude $F({\boldsymbol{r}})$
throughout the junction.  
We point out that the junctions considered here are very wide. They correspond to
the widest case shown in Fig.~\ref{fig:diffwidth}(c)
where the proximity gap is extremely small.
This weak proximity gap is not favorable to topological superconductivity.
This figure should make it clear, however, that even though the channel is wide, the existence of a FFLO phase demonstrates
that
the channel should
be viewed as a proximitized superconductor,
rather than
as a strictly ``normal" region.
Shown in the upper panel of Fig.~\ref{fig:gap2DSOC} are the pair amplitudes for three different values of SOC strength.
The lower panel of Fig.~\ref{fig:gap2DSOC} presents linecuts of this pair amplitude along the $y$ direction at different values of $z$. As can be seen from the figure, the oscillations of the pair amplitude are confined to the
2DEG channel; this oscillation can manifest as the oscillation in the critical current as a function of an in-plane Zeeman field as observed in recent experiments~\cite{hart2017controlled}. The frequency of these oscillations scales appropriately with both
the applied in-plane Zeeman field and the SOC strength.

In Appendix~\ref{sec:appendix} we show that the FFLO state is also present for the case of 1D Rashba SOC. There we also illustrate how the same behavior can be found in the effective models where the superconducting hosts have been ``integrated out".

In general, the amplitude of the FFLO oscillation decreases with increasing temperature~\cite{Heikkinen2013Finite} as temperature weakens FFLO pairing. Since
the typical experimental temperature (from 0.5 K down to 31 mK~\cite{Ren2018Topological})
is well below the superconducting critical temperature ($T_c$ of an Al film is 1.2-1.6 K~\cite{Ren2018Topological}), the FFLO order should be experimentally observable, as reported in recent work
on Al-proximitized HgTe quantum wells~\cite{hart2017controlled}. We note that since the proximitization strength decreases with increasing junction width, the FFLO oscillation amplitude decays towards the middle of the junction (away from the superconductor). This implies that the FFLO phase of a narrower
junction is associated with a larger proximitized gap in the middle of the junction.

The presence of nonmagnetic disorder will decrease the amplitude and period of the FFLO oscillations. This is because the associated scattering involves an averaging of the effective magnitude of the magnetic field over all directions (from a minimum value of 0 to a maximum value of $E_Z$), yielding a shorter oscillation period~\cite{Demler1997Superconducting}. Magnetic disorder, on the other hand, leads to
a decrease in the characteristic decay length and an increase in the period of oscillations~\cite{Buzdin2005Proximity,Buzdin1985Surface}.

\section{Conclusions}\label{sec:conclusion}
While heterostructures that involve proximitization appear to be important for achieving topological
superconductivity,
the major components required to achieve this phase are in many ways inimical to the proximitization
process. These involve Zeeman fields, spin-orbit
coupling which can lead to band-structure
mismatches and substantial chemical potential discontinuities between the parent superconductors
and the proximitized (often semiconducting) medium.
Nevertheless, experiments 
~\cite{Ren2018Topological,Antonio2018Evidence}
seem to be demonstrating success.
Although
theoretically we might expect this proximitization to be a rather delicate and fragile process, nevertheless, we are
able to show that there are clear indications of well-established
topological superconductivity.
The figures throughout this paper illustrate this situation. We stress that in our Josephson-junction configuration
the proximity is more remote compared to that in the 
conventional nanowire configuration of Fig.~\ref{fig:fig1}(b). 

Because we have focused on the proximitization process itself, in this paper we were able to
consider how to maximize the proximity 
gap $\Delta^{\mathrm{prox}}$ both by varying geometry as well as materials 
parameters. This particular parameter $\Delta^{\mathrm{prox}}$ is understood to be computed in the absence of Zeeman 
field or phase difference. It nevertheless sets the scale for the energy 
gap in the topological phase, $E_{\mathrm{gap}}$, and thereby for the 
stability of Majorana zero modes.

Figure~\ref{fig:gapthickness} presents a summary of our major findings. 
One should aim for junctions with very thin 2DEG regions and narrow 
channels between the host superconductors. Additionally, there is a 
delicate competition between the chemical potential differences of the 
2DEG and the superconductors ($\delta \mu$), and the Rashba SOC strength. While a larger $\delta \mu$ serves to compensate for 
deleterious effects of SOC, it cannot be too big. Indeed, 
Fig.~\ref{fig:phasediagram2}(b) shows that one major knob of the 
Josephson junction architecture (which is the ability to tune the phase 
difference to $\pi$ and thereby require very small Zeeman fields to 
access topological phases) is undermined if $\delta \mu$ is too large.

Finally, by plotting the pair amplitude itself,
we have provided in this paper very direct evidence for the 
elusive FFLO phase. It is not necessarily to be associated with 
topological physics, but it has some of the same requirements. We show 
how the presence of Zeeman fields together with SOC and (remote) proximity effect 
stabilize this state which exists entirely inside the 2DEG channel, much 
as in recent experiments \cite{hart2017controlled}.

\acknowledgements
We thank Erez Berg for helpful conversations. This work was supported by NSF-DMR-MRSEC 1420709. C.-T.W. is supported by the MOST Grant No. 106-2112-M-009-001-MY2. We acknowledge the University of Chicago Research Computing Center for support of this work.
\bibliography{paper_revised2}
\onecolumngrid
\appendix

\numberwithin{equation}{section}
\numberwithin{figure}{section}
\section{FFLO phases}\label{sec:appendix}

\begin{figure}
\includegraphics[width=0.5\linewidth]
{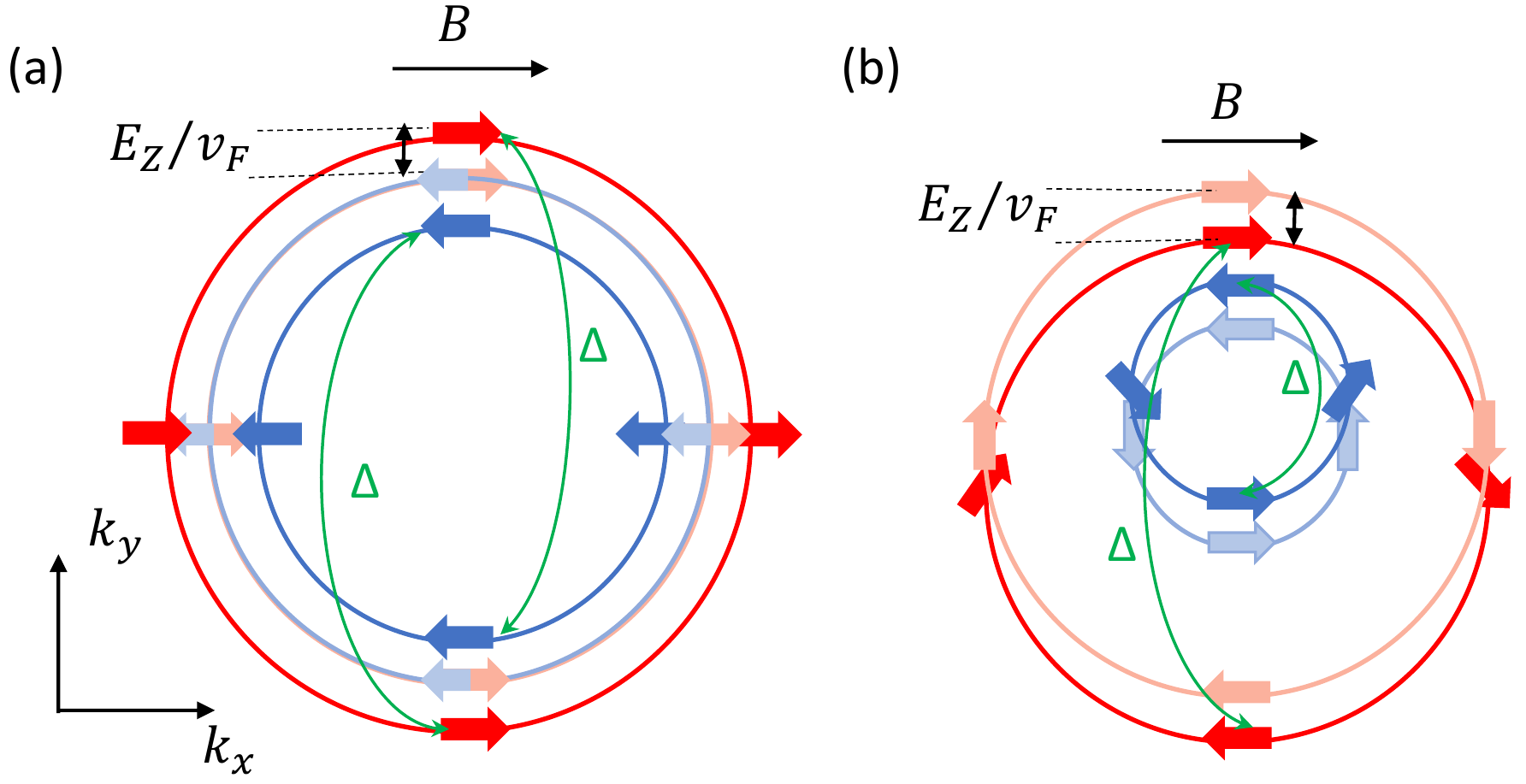}
\caption{The change of Fermi surfaces of a 2DEG due to an in-plane magnetic field $B$ along the junction ($x$ direction) for the case of (a) zero SOC and (b) finite SOC strength. The Fermi surfaces in the absence and presence of $B$ are represented by light and dark colors, respectively. (a) In the absence of a Zeeman field, the Fermi surfaces of a 2DEG without SOC are doubly degenerate. When an in-plane Zeeman field is applied, the Fermi surfaces of the up  and down spins enlarge and shrink radially in momentum by $E_Z/v_F$ while keeping the two Fermi surfaces concentric. The superconducting term $\Delta$ pairs up electrons with opposite spin from different Fermi surfaces. (b) The 2D Rashba SOC causes a clockwise and anticlockwise spin orientation (represented by red and blue arrows, respectively). The applied in-plane Zeeman field along the $x$ direction shifts the inner and outer Fermi surfaces in the opposite direction along $k_y$ by $E_Z/v_F$. The superconductivity term $\Delta$ pairs up electrons with opposite spin from the same Fermi surface.}
\label{fig:Fermisurfaces}
\end{figure}

We begin by studying the mechanism for the formation of the FFLO phases. 
In the absence of SOC, the Fermi surfaces of up and down spins always form concentric circles as shown in Fig.~\ref{fig:Fermisurfaces}(a). For zero Zeeman fields, the superconducting pairing occurs between electrons carrying opposite spin with opposite momentum ($\boldsymbol{k}\uparrow$ and $-\boldsymbol{k}\downarrow$) on the Fermi surface where the 
Cooper pair has a zero center of mass momentum. If an in-plane magnetic field is applied to a system with no SOC, the Zeeman field enlarges and shrinks the Fermi surfaces radially in momentum by $E_Z/v_F$ for the up and down spins, respectively, while keeping the two Fermi surfaces concentric. The pairing now occurs between the up- and down-spin electrons with different Fermi momenta, i.e., $\boldsymbol{k} + \boldsymbol{q}/2$ and $-\boldsymbol{k}+\boldsymbol{q}/2$ where $q = 2E_Z/v_F$, so that
the Cooper pairs have a net center of mass momentum of $\boldsymbol{q}$. When the applied in-plane Zeeman field is sufficiently strong, spatial symmetry needs to be broken in order to lower the ground state energy which results in the FFLO state. However, because of the Pauli depairing, this FFLO state only survives in a narrow parameter regime. This depairing effect in strong Zeeman fields can be mitigated by using the SOC, which allows both singlet and triplet pairings, since the triplet pairing is not sensitive to the depairing effect.

In the presence of Rashba SOC, the Hamiltonian of a 2DEG without a Zeeman field [Eq.~\eqref{eq:BdG}] is invariant when the spin and momentum are rotated simultaneously in the $x$-$y$ plane, i.e.,
\begin{align}
\left(\begin{matrix}  -k'_y \\ k'_x \end{matrix}\right) = \mathcal{R} \left(\begin{matrix}  -k_y \\k_x \end{matrix}\right), \hspace{0.5 cm} \left(\begin{matrix}  \sigma'_x \\ \sigma'_y \end{matrix}\right) = \mathcal{R} \left(\begin{matrix}  \sigma_x \\ \sigma_y \end{matrix}\right),
\end{align}
where 
\begin{align}
\mathcal{R} = \left(\begin{matrix}  \cos \theta & \sin \theta  \\ - \sin\theta & \cos\theta  \end{matrix}\right)
\end{align}
is the rotation operator in the $x$-$y$ plane. 

Note that the Hamiltonian still respects this rotational symmetry even in the presence of an out-of-plane Zeeman field (along the $z$ direction). However, the application of an in-plane Zeeman field $E_Z$ along the junction, i.e., along the $x$ direction, breaks this rotational symmetry. The energy spectrum of the electron in the presence of the in-plane Zeeman field $E_Z$ is given by
\begin{equation}
E = k_x^2 + k_y^2 - \mu + \frac{\alpha^2}{4} \pm \sqrt{\alpha^2k_x^2 + (E_Z - \alpha k_y)^2}.
\end{equation}
which breaks the rotational symmetry.

In the limit where $E_Z \ll \alpha k_F \ll \mu$, the two Fermi surfaces are shifted in the direction perpendicular to the Zeeman field direction (along $k_y$) by $q = 2E_Z/v_F$ as shown in Fig.~\ref{fig:Fermisurfaces}(b). The pairing in this case occurs between up and down spins belonging to the same Fermi surface resulting also in Cooper pairs having a net momentum of $\boldsymbol{q}$. Thus the wave function of the Cooper pair can be written as $\cos(qy)|S\rangle + \sin(qy)|T\rangle$, where  $|S\rangle = \left|\uparrow\downarrow\right\rangle - \left|\downarrow\uparrow\right\rangle$  and $|T\rangle = \left|\uparrow\downarrow\right\rangle +\left|\downarrow\uparrow \right\rangle$ are the singlet and triplet pairing wave functions, respectively. So, the presence of SOC stabilizes the FFLO phase as the SOC lifts the spin degeneracy and shifts the Fermi surface in such a way that the resulting Cooper pair has a finite center of momentum~\cite{Zheng2013Route,zheng2014fflo}. 

\begin{figure*}
\includegraphics[width=0.8\linewidth]
{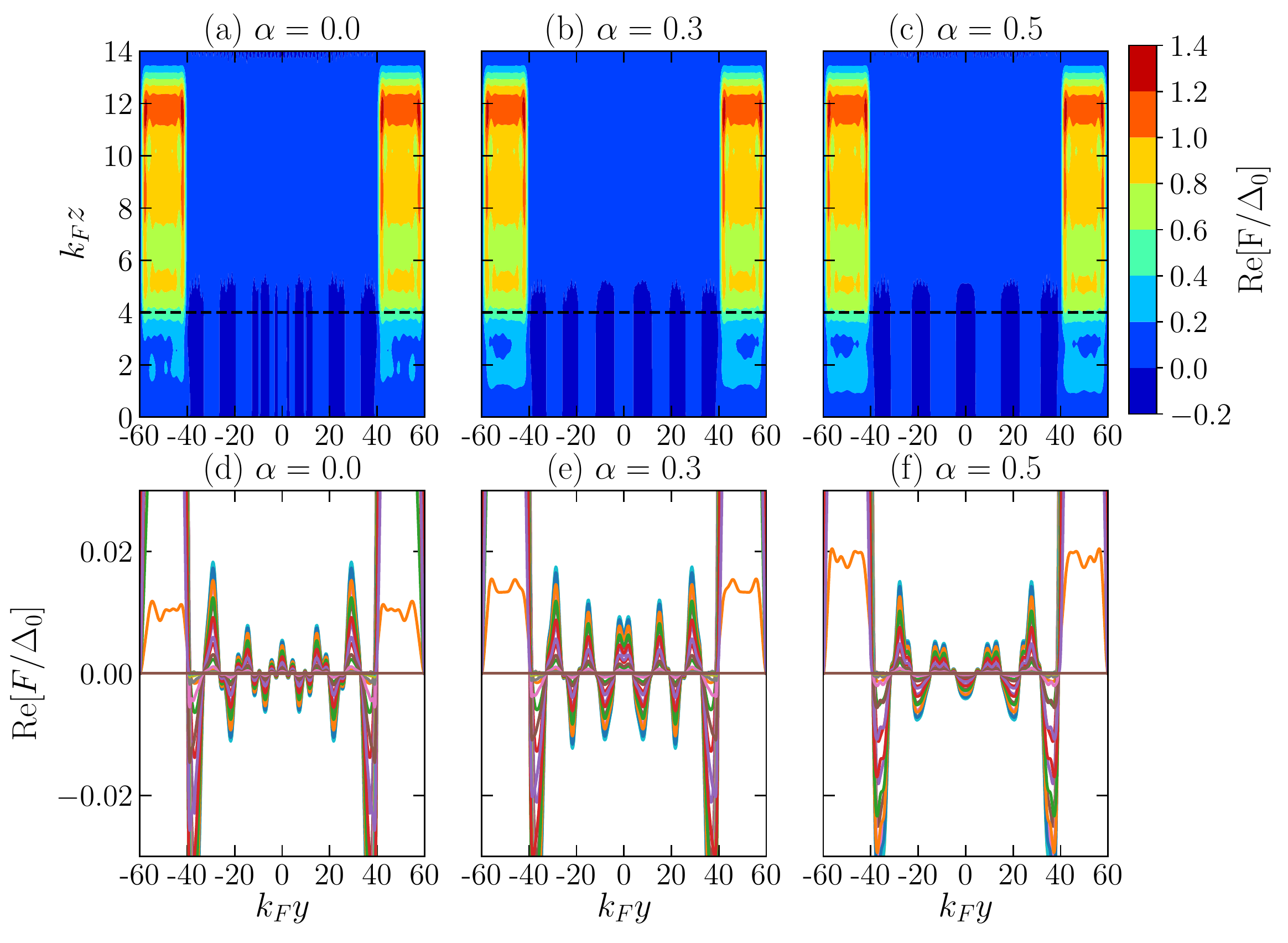}
\caption{Evidence for FFLO. Profile of the real part of the pair amplitude of a planar Josephson junction with a 1D Rashba SOC ($\alpha\partial_y\sigma_x$) of different strengths: $\alpha = 0$ (left panel), $\alpha = 0.3$ (middle panel), and $\alpha = 0.5$ (right panel). Upper panel shows the color plots of the real part of the pair amplitude Re[$F_0/\Delta_0$], and lower panel shows the linecuts of the pair amplitude along the $y$ direction at different values of $z$.  As shown in the lower panel, the pair amplitude oscillates along the $y$ direction with a characteristic oscillation length  $\lambda = E_Z/(2v_F)$ that decreases with increasing $\alpha$ (as $v_F$ increases with increasing $\alpha$). The black dashed lines in the upper panel denote the boundaries between the superconductors and the 2DEG. The parameters used are $\mu_{\mathrm{S}} = 1$, $\mu_{\mathrm{2DEG}} = 1$, $E_{Z,J} = E_{Z,L} = 0.2$, $\Delta_{0}$ = 0.3, $\phi =0$, $W_{\mathrm{SC}}$ = 20/$k_{F}$, $W$ = 80/$k_{F}$,  $D_{\mathrm{SC}}$ = 10/$k_{F}$, and $D_{\mathrm{2DEG}}$ = 4/$k_{F}$.}
\label{fig:gap1DSOC}
\end{figure*}

In the main text we have shown how the FFLO phase appears in a proximitized junction in the presence of an in-plane Zeeman field and a conventional (2D) Rashba SOC. 
In this appendix we show that our findings are quite robust, appearing also for a 1D Rashba SOC as well as in the effective model. We self-consistently solve the BdG equations to obtain
the pair amplitude [as given by Eq.~\eqref{eq:Deltaself} of the main text]:
\begin{align}\label{eq:Deltaself1}
F(y,z) =  \int dk_x \sum_{E_{nk_x}<\omega_D} \left[u_{nk_x\uparrow}v_{nk_x\downarrow}^*-u_{nk_x\downarrow}v_{nk_x\uparrow}^*\right]\tanh\left(\frac{E_{nk_x}}{2T}\right).
\end{align}

Figure~\ref{fig:gap1DSOC} shows the pair amplitude $F(y,z)$  for a 2DEG with a 1D Rashba spin-orbit-coupling $\alpha\partial_y\sigma_x$. The pair amplitudes are calculated for different SOC strengths. As for the case of 2D Rashba spin-orbit-coupled electron gas, here we also find an oscillation of the pair amplitude within the junction channel and
with the oscillation length scale given by $\lambda = E_Z/(2 v_F)$ which increases with increasing Zeeman field $E_Z$ and decreases with increasing $\alpha$ (as $v_F$ increases with increasing $\alpha$). This is indicative of the FFLO phases formed in the presence of 
an applied in-plane magnetic field along the junction. We note that the Hamiltonian of a 2DEG with a 1D Rashba SOC can be mapped by a gauge transformation into the Hamiltonian of a conical Holmium magnet~(Ref.~\cite{Wu2018Quantum}) or coupled nanowires~(Ref.~\cite{Wang2014Fate}) which are also platforms for topological superconductors.

\begin{figure*}
\includegraphics[width=0.8\linewidth]
{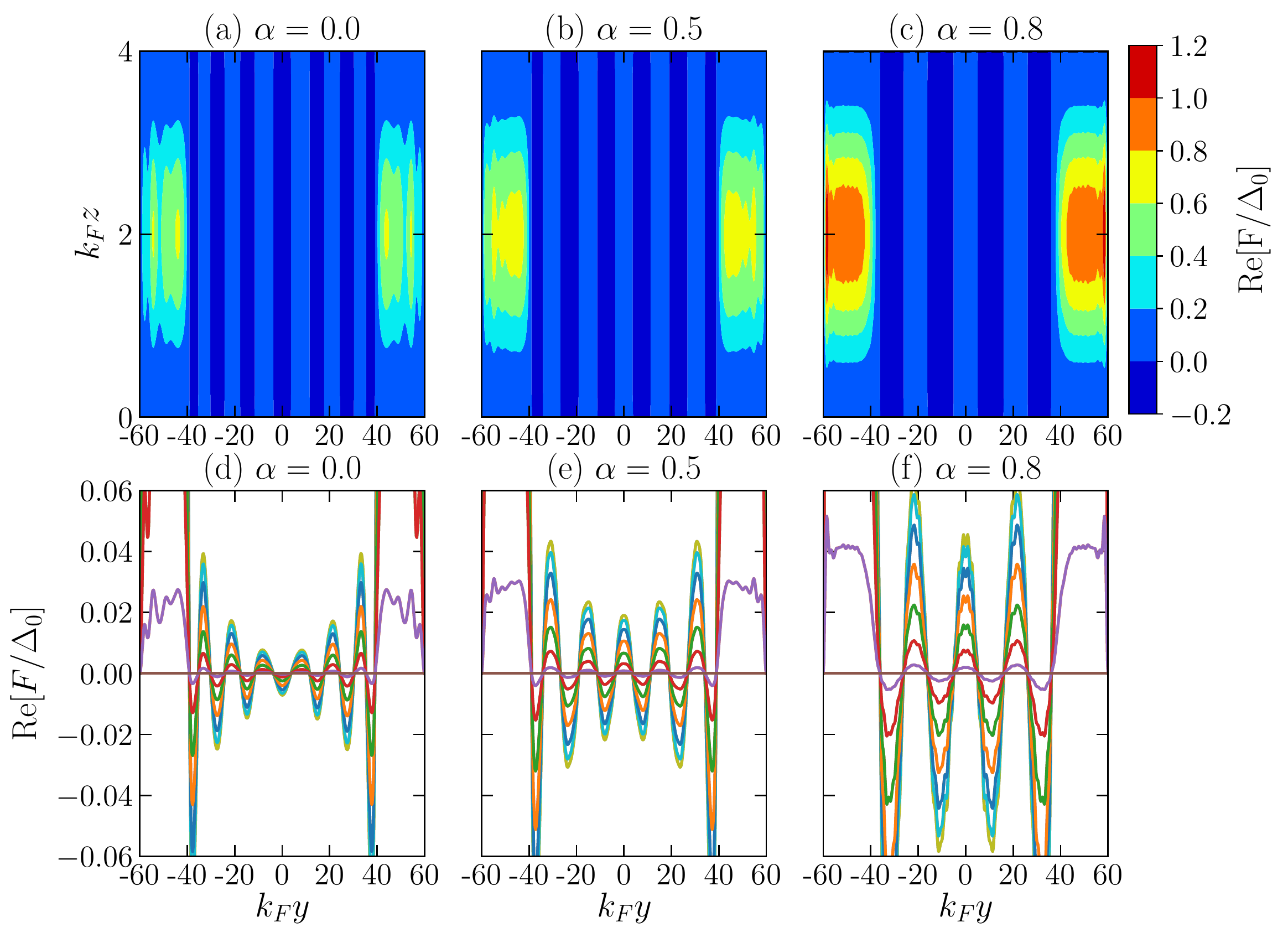}
\caption{Evidence for FFLO. Profile of the real part of the pair amplitude for the effective model of a planar Josephson junction [Eq.~\eqref{eq:HSM} of the main text]. The pair amplitudes are calculated for a 2D Rashba SOC of different strengths: $\alpha = 0$ (left panel), $\alpha = 0.5$ (middle panel), and $\alpha = 0.8$ (right panel). Upper panel shows the color plots of the real part of the pair amplitude Re[$F_0/\Delta_0$], and lower panel shows the linecuts of the pair amplitude along the $y$ direction at different values of $z$. As shown in the lower panel, the pair amplitude oscillates along the $y$ direction with a characteristic oscillation length  $\lambda = E_Z/(2v_F)$ that decreases with increasing $\alpha$ (as $v_F$ increases with increasing $\alpha$).  The parameters used are $\mu_{\mathrm{S}} = 1$, $\mu_{\mathrm{2DEG}} = 1$, $E_{Z,J} = E_{Z,L} = 0.3$, $\Delta_{0}$ = 0.3, $\phi =0$, $W_{\mathrm{SC}}$ = 20/$k_{F}$, $W$ = 80/$k_{F}$, and $D_{\mathrm{2DEG}}$ = 4/$k_{F}$.}
\label{fig:gaptoy}
\end{figure*}

Finally, Fig.~\ref{fig:gaptoy} shows the pair amplitude $F(y,z)$ for the effective model [Eq.~\eqref{eq:HSM}] of a planar Josephson junction with a 2D Rashba SOC. As shown in the figure, in the presence of an in-plane magnetic field, the pair amplitude $F(\boldsymbol{r})$ oscillates inside the junction channel with an oscillation length which decreases with increasing  SOC strength. Again, the oscillation is 
consistent with the formation of an FFLO phase in the presence of an in-plane magnetic field.

\end{document}